\documentclass[aps,prd,twocolumn,superscriptaddress,amssymb,nofootinbib,floatfix]{revtex4}
\usepackage{fancyhdr}
\usepackage{fancybox}
\usepackage{graphicx}
\usepackage{color}
\usepackage{soul}
\usepackage{latexsym}

\def\cvisq{C_V^2(\hi)}

\def\tauptaum{\tau^+\tau^-}

\def\cbb{C_{eff}^{2b}}

\def\mh{m_h}
\def\hi{h_1}
\def\hii{h_2}
\def\ai{a_1}
\def\mhi{m_{h_1}}
\def\mhii{m_{h_2}}
\def\mai{m_{a_1}}
\def\mueff{\mu_{\rm eff}}

\def\lam{\lambda}
\def\kap{\kappa}
\def\alam{A_\lambda}

\def\akap{A_\kappa}

\def\mh{m_h}

\def\mhusq{m_{H_u}^2}
\def\mhdsq{m_{H_d}^2}
\def\mssq{m_S^2}

\def\h{h}
\def\mh{m_{h}}

\def\hbar{\overline h}
\def\lam{\lambda}

\def\mz{m_Z}

\def\hi{h_i^0}
\def\mhi{m_{\hi}}

\def\h{h}
\def\mh{m_{\h}}

\def\lam{\lambda}

\def\cosb{\cos\beta}

\def\sinb{\sin\beta}

\def\nn{\nonumber}

\def\wtil{\widetilde}

\def\tauptaum{\tau^+\tau^-}

\def\lsim{\mathrel{\raise.3ex\hbox{$<$\kern-.75em\lower1ex\hbox{$\sim$}}}}
\def\gsim{\mathrel{\raise.3ex\hbox{$>$\kern-.75em\lower1ex\hbox{$\sim$}}}}
\def\ifmath#1{\relax\ifmmode #1\else $#1$\fi}
\def\half{\ifmath{{\textstyle{1 \over 2}}}}

\def\third{\ifmath{{\textstyle{1 \over 3}}}}

\def\vev#1{\langle #1 \rangle}
\def\lam{\lambda}

\def\mhi{m_{h_1}}

\def\eg{{\it e.g.}}

\def\stop{\wt t}

\def\mstopbar{\overline m_{\stop}}

\def\msusy{m_{\rm SUSY}}

\def\susy{{\rm SUSY}}

\def\eg{{\it e.g.}}

\def\hsm{h_{\rm SM}}

\def\hl{h}
\def\hh{H}
\def\ha{A}

\def\mhl{m_{\hl}}
\def\mhh{m_{\hh}}
\def\mha{m_{\ha}}

\def\tanb{\tan\beta}

\def\mb{m_b}
\def\mz{m_Z}

\def\mgut{M_U}

\def\wt{\widetilde}

\def\MPL #1 #2 #3 {{\sl Mod.~Phys.~Lett.}~{\bf#1} (#3) #2}
\def\NPB #1 #2 #3 {{\sl Nucl.~Phys.}~{\bf #1} (#3) #2}
\def\PLB #1 #2 #3 {{\sl Phys.~Lett.}~{\bf #1} (#3) #2}
\def\PR #1 #2 #3 {{\sl Phys.~Rep.}~{\bf#1} (#3) #2}
\def\PRD #1 #2 #3 {{\sl Phys.~Rev.}~{\bf #1} (#3) #2}
\def\PRL #1 #2 #3 {{\sl Phys.~Rev.~Lett.}~{\bf#1} (#3) #2}
\def\RMP #1 #2 #3 {{\sl Rev.~Mod.~Phys.}~{\bf#1} (#3) #2}
\def\ZPC #1 #2 #3 {{\sl Z.~Phys.}~{\bf #1} (#3) #2}
\def\IJMP #1 #2 #3 {{\sl Int.~J.~Mod.~Phys.}~{\bf#1} (#3) #2}
\def\NIM #1 #2 #3 {{\sl Nucl.~Inst.~and~Meth.}~{\bf#1} {#3} #2}
\def\lam{\lambda}
\def\br{B}
\def\tauptaum{\tau^+\tau^-}

\def\gam{\gamma}

\def\anti{\overline}
\def\epem{e^+e^-}

\def\ie{{\it i.e.}}
\def\eg{{\it e.g.}}

\def\anti{\overline}

\def\ai{a_1}
\def\aii{a_2}
\def\mai{m_{\ai}}

\def\gev{~{\rm GeV}}
\def\tev{~{\rm TeV}}

\def\mb{m_b}

\def\hi{\h_1}
\def\hii{\h_2}
\def\hiii{\h_3}
\def\mhi{m_{\hi}}
\def\mhii{m_{\hii}}

\newcommand{\nc}{\newcommand}
\nc{\beq}{\begin{equation}}   \nc{\eeq}{\end{equation}}
\nc{\bea}{\begin{eqnarray}}   \nc{\eea}{\end{eqnarray}}
\nc{\baa}{\begin{array}}      \nc{\eaa}{\end{array}}
\nc{\bit}{\begin{itemize}}    \nc{\eit}{\end{itemize}}
\nc{\ben}{\begin{enumerate}}  \nc{\een}{\end{enumerate}}
\nc{\bce}{\begin{center}}     \nc{\ece}{\end{center}}
\def\beqa{\begin{eqnarray}}
\def\eeqa{\end{eqnarray}}
\def\bed{\begin{description}}
\def\eed{\end{description}}

\def\mhi{m_{h_1}}

\def\eg{{\it e.g.}}

\def\half{\frac{1}{2}\,}
\def\third{\frac{1}{3}\,}

\def\tanb{\tan\beta}

\def\simle{%  ``less than about'' symbol
    \mathrel{\rlap{\raise 0.511ex 
        \hbox{$<$}}{\lower 0.511ex \hbox{$\sim$}}}}

\def\slashchar#1{\setbox0=\hbox{$#1$}           % set a box for #1
   \dimen0=\wd0                                 % and get its size
   \setbox1=\hbox{/} \dimen1=\wd1               % get size of /
   \ifdim\dimen0>\dimen1                        % #1 is bigger
      \rlap{\hbox to \dimen0{\hfil/\hfil}}      % so center / in box
      #1                                        % and print #1
   \else                                        % / is bigger
      \rlap{\hbox to \dimen1{\hfil$#1$\hfil}}   % so center #1
      /                                         % and print /
   \fi}

\def\lam{\lambda}

\def\ls#1{\ifmath{_{\lower1.5pt\hbox{$\scriptstyle #1$}}}}
\def\lss#1{\ifmath{^{\,\lower2.5pt\hbox{$\scriptstyle #1$}}}}

\def\nn{\nonumber}
\tighten \preprint{UCD-HEP-???}
\begin{document}

\title{\Large\bf A Comparison of Mixed-Higgs Scenarios In the NMSSM and the MSSM}
\author{
Radovan Derm\' \i \v sek}
\address{
School of Natural Sciences, Institute for Advanced Study, Princeton,
NJ 08540}
%\footnote{dermisek@physics.ucdavis.edu} 
%\footnote{gunion@physics.ucdavis.edu}
\author{John F. Gunion}
\address{
Department of Physics, University of California at Davis, Davis, CA 95616} 
\begin{abstract} 
  We study scenarios in the minimal and next-to minimal supersymmetric
  models in which the lightest CP-even Higgs boson can have mass below
  the $114\gev$ standard model LEP limit by virtue of reduced $ZZ$ coupling
  due to substantial mixing among the Higgs bosons. We
  pay particular attention to the size of corrections from
  superpartners needed for these scenarios to be viable and point to
  boundary conditions at large scales which lead to these scenarios
  while at the same time keeping electroweak fine tuning modest in
  size.
%In the
%  minimal supersymmetric model it is found that the size of
%  corrections needed is possibly smaller (in a limited region of $\tan
%  \beta$) but typically comparable to the size of corrections required
%  in the usual decoupled scenario in which the lightest CP-even Higgs
%  boson is standard model like and heavier than 114.4 GeV.
%  Furthermore, if the lightest CP-even Higgs is required to explain
%  the LEP excess of Higgs like events at about 98 GeV the size of
%  corrections from superpartners is identical in the MSSM and NMSSM.
  We find that naturalness of electroweak symmetry breaking in the
  mixed-Higgs scenarios of both
  models points to the same region of soft supersymmetry breaking
  terms, namely those leading to large mixing in the stop sector at
  the electroweak scale, especially if we also require that the
  lightest CP-even Higgs explains the Higgs-like LEP events at $\sim
  98\gev$. 

\end{abstract}
\maketitle
%[Key words: Electroweak Symmetry Breaking,
%Supersymmetry Breaking, Color/Charge Breaking]
%\pacs{}
\thispagestyle{empty}

\section{Introduction}

Supersymmetry  cures the naturalness / hierarchy
problem associated with the quadratically divergent 1-loop corrections
to the Higgs boson mass 
via the introduction of superpartners for each SM particle.
So long as the superpartners have mass somewhat below $1\tev$,
the cancellation is not particularly extreme and the hierarchy /
naturalness problem associated with the quadratic divergences is
ameliorated.  However, there remains the question of how finely the
GUT-scale parameters must be adjusted in order to get appropriate
electroweak symmetry breaking, that is to say correctly predict the
observed value of $\mz$.  LEP limits on a
SM-like Higgs boson play a crucial role here.

Supersymmetric models most naturally predict that the lightest Higgs
boson, generically $h$, has couplings to $ZZ$ and $f\anti f$ pairs of
SM strength (such an $h$ is termed 'SM-like') and that it has a mass
closely correlated to $\mz$, typically lying in the range $\lsim
105\gev$ for stop masses $\lsim 500\gev$, with an upper bound, for
example, of $\lsim 135\gev$ in the MSSM for stop masses $\sim 1\tev$
and large stop mixing.  If the stop masses are large, the predicted
value of $\mz$ is very sensitive to the GUT scale parameters. Such
sensitivity is termed `fine tuning'.  Models with minimal fine tuning
provide a much more natural explanation of the $Z$ mass than those
with a high level of fine tuning. The degree of fine tuning required
is thus quite closely related to the constraints on a SM-like $h$, and
these in turn depend on how it decays.

The SM and the MSSM predict that $h\to b\anti b$ decays are dominant
and LEP has placed strong constraints on $\epem\to Zh\to Z b\anti b$.  The
limits on the effective coupling
\beq
\cbb\equiv \left[{g_{ZZh}^2 \over g_{ZZ\hsm}^2}\right]\br(h\to b\anti b)
\eeq 
are such that $\mh<114\gev$ is excluded for a SM-like $h$ that decays
primarily to $b\anti b$.  For $\msusy\lsim 1\tev$, most of
CP-conserving MSSM parameter space is ruled out by this LEP limit.
There are three surviving parts of MSSM parameter space. The first
such part is characterized by at least one large stop mass at or above
a $\tev$ at scale $\mz$.  In this case, it is always the case that to
predict the observed $\mz$ requires very careful adjustment, \ie\ fine
tuning, of the GUT-scale parameters (either the Higgs mass-squared or
$\mu^2$) with accuracies better than 1\% (the smaller the percentage
accuracy required, the more fine-tuned is the model). The second part
of MSSM parameter space that is consistent with LEP limits by virtue
of having $\mh>114\gev$ is that with large mixing in the stop sector
(\ie\ large $|A_t|/\mstopbar$), where $\mstopbar\equiv
[\half(m_{\tilde t_1}^2+m_{\tilde t_2}^2)]^{1/2} \gsim 300\gev$ and
$A_t\lsim -500\gev$ (at scale $\mz$).  This was explored in our
previous paper, where we found that fine tuning could be improved to
about the $3\%$ level. The third part of MSSM parameter space
consistent with LEP limits is that where strong mixing between the two
CP-even scalars of the model takes place, as arises when the CP-odd
$\ha$ has mass $\mha\sim 100\gev$. In this region, the lightest
CP-even Higgs has mass somewhat below the SM LEP limit of $114\gev$,
as allowed by virtue of reduced $ZZ$ coupling due to the mixing, and
the heavier CP-even Higgs boson has mass slightly above this value.  A
mass $\mhh>114\gev$ is achieved by virtue of both the effects of Higgs
mixing and large radiative corrections from the stop sector. However,
because of the Higgs mixing the latter stop sector corrections need
not be as large as in the parts of parameter space for which
$\mh>114\gev$.  In the first part of this paper, we explore this third
sector of MSSM parameter space in detail. It is characterized by the
extension of the first two regions to smaller stop masses or to
smaller $|A_t|/\mstopbar$. The first region is extended to stop masses
of $\mstopbar\gsim 600\gev$, leading to fine tuning of order $2\%$.
The second region is extended to somewhat smaller $\mstopbar$ and
significantly smaller ratio of $|A_t|/\mstopbar$, for which we find
that the GUT scale parameters must be chosen with an accuracy of at
least $6.5\%$. This is a significant decrease of fine tuning relative
to the other cases.

In the second part of the paper, we consider mixed-Higgs scenarios in
the Next-to-Minimal Supersymmetric Model (NMSSM) yielding a lightest
Higgs boson $\hi$ with $\mhi<114\gev$ that escapes LEP limits by
virtue of Higgs mixing yielding reduced $ZZ\hi$ coupling. Two basic
types of Higgs mixing can yield reduced $ZZ\hi$ coupling while keeping
fine tuning to a not too unacceptable level: i) mixing of the two
doublet Higgs fields analogous to MSSM mixed-Higgs scenarios; and ii)
mixing of the doublet Higgs fields with the singlet Higgs field.  In NMSSM
case i), our scans have found parameters yielding MSSM-like
mixed-Higgs scenarios with the same level of fine tuning
as in mixed-Higgs MSSM scenarios, \ie\ $\sim 6.5\%$.
In NMSSM case ii), we find it is also possible to reduce the 
GUT-scale parameter tuning
required for correct EWSB to the level of $\sim 6.5\%$. We will
present details of Higgs masses and GUT-scale parameters associated
with these scenarios.

Although not the focus of this paper, the NMSSM mixed-Higgs scenarios
should always be thought of in comparison to the very natural $\sim
17\%$ fine-tuning scenarios where the $\hi$ is very SM-like and has
mass $\mhi\sim 100\gev$. In this case, see
\cite{Dermisek:2005ar,Dermisek:2005gg,Dermisek:2007yt}, the $\hi$
evades LEP limits by virtue of its primary decay being $\hi\to\ai\ai$
where $\mai<2\mb$ so that the rate for $\epem\to Z\hi\to Z+b's$ (where
$b's$ refers to any final state with 2 or more $b's$) is
small.\footnote{The importance of Higgs to Higgs decays was first made apparent
  in \cite{ourfirstpaper} and \cite{Gunion:1996fb}. Further
  experimental implications of such decays were explored in Refs.~\cite{Ellwanger:2001iw,Ellwanger:2005uu}.}  The
attractiveness of this scenario is not only that it is not at all
fine-tuned, but also: i) a SM-like Higgs with mass near $100\gev$ is
strongly preferred by precision electroweak measurements; and ii)
these scenarios with large $\br(\hi\to\ai\ai)$ typically
predict~\cite{Dermisek:2005gg} an excess in the $\epem \to Z+b's$
quite consistent with well-known $2.3\sigma$ excess in the LEP data
for $M_{b's}\sim 98\gev$~\cite{oldleplimits}.  Meanwhile, there are no
current limits on the $Z\hi\to Z\ai\ai\to Z\tauptaum\tauptaum$ final
state for $\mhi\gsim 87\gev$ \cite{newleplimits}. And limits in the
case of $\ai\to jets$ run out at still lower $\mhi$.

In order to quantify fine tuning, we employ the measure
\beq 
F\equiv {\rm Max}_p F_p \equiv {\rm Max}_p\left|{d\log \mz\over d\log
    p}\right|\,, 
\label{fdef}
\eeq
where the parameters $p$ comprise all GUT-scale soft-SUSY-breaking
parameters. Above, we used $F^{-1}$ in percent to
express the degree of fine tuning.  The larger the fine-tuning $F$, the
more finely the most sensitive GUT-scale parameter must be
tuned (adjusted) as a percentage of its nominal value.

While there are many earlier papers that have considered mixed-Higgs
scenarios in the context of both the 
MSSM~\cite{Sopczak,Carena:2000ks,Kane:2004tk,Drees:2005jg,Belyaev:2006rf,Kim:2006mb,Essig:2007vq} and 
NMSSM (or other singlet extensions of the MSSM)~\cite{BasteroGil:2000bw,Barger:2006dh,Barger:2006sk},
most did not consider the fine tuning issue.
Only a few papers~\cite{BasteroGil:2000bw,Kim:2006mb,Essig:2007vq} 
have studied the correlations between fine tuning
and Higgs mixing.  This paper will extend these latter studies, fully
exploring all of parameter space.

%%%%%%%%%%%%%%%%%%%%%%%%%%%%%%%%%%%%%%%%%%%%%%%%
\section{MSSM}
%%%%%%%%%%%%%%%%%%%%%%%%%%%%%%%%%%%%%%%%%%%%%%%%

In the MSSM, the CP-even Higgs mass-squared matrix in the basis ($H_d$, $H_u$) is given as:
\begin{equation}
M \simeq \left(    \begin{array}{cc}
                    m_A^2 s^2_\beta + m_Z^2 c^2_\beta & -(m_A^2 + m_Z^2 ) s_\beta c_\beta \\
                    -(m_A^2 + m_Z^2 )s_\beta c_\beta & m_Z^2 s^2_\beta + m_A^2 c^2_\beta + \Delta
                    \end{array} \right),
\end{equation}
where $m_A$ is the mass of the CP odd Higgs boson, $m_Z$ is the mass
of the Z boson, $\tan \beta = v_u/v_d$ is the ratio of the vacuum
expectation values of the two Higgs doublets and we use the shorthand
notation $c_\beta = \cos \beta$ and $s_\beta = \sin \beta$. Finally,
$\Delta$ is the SUSY correction to the 2-2 element of $M$ which is
dominated by the contributions from stop loops and thus depends on
stop masses and the mixing in the
stop sector. It is the size of this correction which is relevant for
the discussion of fine tuning of electroweak symmetry breaking.

The mass eigenstates are defined as follows:
\begin{equation}
\left(  \begin{array}{c} H \\ h \end{array} \right)
                  = \left( \begin{array}{cc}
                         c_\alpha &  s_\alpha \\
                          -s_\alpha & c_\alpha
                    \end{array} \right) 
                    \left(  \begin{array}{c} H_d \\ H_u \end{array} \right),
\end{equation}
and the coupling squared of the lighter CP-even Higgs boson to $ZZ$ divided by the standard model value is given as:
\begin{equation}
\xi^2 = \frac{g^2_{ZZh}}{g^2_{ZZh_{SM}}} =  \sin^2 (\beta - \alpha).
\end{equation}
(Note that in the notation of the NMSSM section of this paper, an
equivalent notation would be $C_V^2(\hl)$ in place of $\xi^2$.)
Introducing a dimensionless quantity:
\begin{equation}
r_\Delta = \frac{\Delta}{m_Z^2}
\end{equation}
and assuming $\tan \beta > {\rm few}$ 
we can rewrite the CP-even Higgs mass-squared matrix as:
\begin{equation}
M \simeq \left(    \begin{array}{cc}
                    m_A^2   & -(m_Z^2 + m_A^2 ) s_\beta c_\beta \\
                    - (m_Z^2 + m_A^2 )  s_\beta c_\beta & m_Z^2 (1 + r_\Delta )
                    \end{array} \right).
\end{equation}
Let us discuss the Higgs sector in two limits:
\begin{itemize}
\item $ m_A \gg m_Z$ -- decoupled Higgs scenario: the lighter CP-even
  Higgs boson originates from $H_u$, $\alpha \simeq 0$, its mass is
  $m_h^2 \simeq m_Z^2 (1 + r_\Delta)$ and it has SM-like $ZZ$
  coupling, $ \xi^2 \simeq 1$.

\item $ m_A^2 < m_Z^2 (1 + r_\Delta )$ -- mixed-Higgs scenario: the
  lighter CP-even Higgs boson originates mainly from $H_d$, $m_h
  \simeq m_A$, and it has reduced $ZZ$ coupling, $\xi^2 \ll 1$. The heavier
  Higgs originates from $H_u$, its mass is $m_H^2 \simeq m_Z^2 (1 +
  r_\Delta)$ and it has SM-like $ZZ$ coupling.
\end{itemize}
Without off diagonal elements (mixing) in the Higgs mass-squared
matrix, both scenarios require exactly the same size for the SUSY
correction, $\Delta$, and thus the same level of EWSB fine tuning,
in spite of the fact that in the mixed-Higgs scenario the lighter
Higgs is well below the LEP limit of 114.4 GeV on the SM Higgs boson.
The reason is that it is the SM-like Higgs (the Higgs with near
maximal $ZZ$ coupling) which has to be pushed above the LEP limit
irrespectively of the fact that it is the heavier of the two.

The off diagonal element in the Higgs mass-squared matrix makes the
heavier eigenvalue heavier and the lighter eigenvalue lighter and
thus, while in the usual decoupled scenario it decreases the mass of
the SM-like Higgs boson, in the mixed-Higgs scenario it increases the
mass of the SM-like Higgs boson. Thus, in the presence of mixing, the
SUSY correction from the stop sector does not have to be as large in
the mixed-Higgs scenario as in the decoupled scenario.\footnote{This
  possibility of increasing the Higgs mass by mixing was suggested as
  a solution to the fine tuning problem of EWSB in
  Ref.~\cite{Kim:2006mb}.}  However, the mixing can be used to
increase the Higgs boson mass only to some extent. First of all, the
mixing term is proportional to $s_\beta c_\beta$ and so for small or
large $\tan \beta$ it is negligible.  For moderate $\tanb$,
$ZZ$ couples almost entirely to $H_u$, but the mixing term in $M$ can
still lead to the $ZZ$ coupling being shared between the
two Higgs mass eigenstates. However, the off-diagonal term cannot be
too large without the light, mainly $H_d$, mass state having a $\xi^2$ 
that exceeds the very strong limits on this quantity for
any Higgs with mass well below the LEP limit.

\begin{figure}[t]
\includegraphics[width=2.5in]{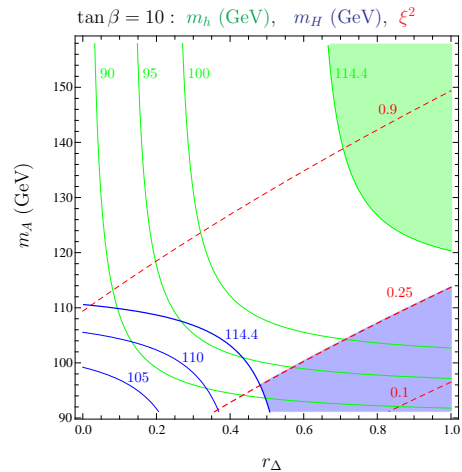}
\caption{
  Contours of constant $m_h$ (green), $m_H$ (blue) and $\xi^2$ (red)
  in the $m_A$ -- $r_\Delta$ plane for $\tan \beta = 10$. The green shaded
  region represents the allowed region for the decoupled scenario (the
  light CP-even Higgs is above the LEP limit on the mass of the SM
  Higgs boson) and the blue shaded region represents the allowed region
  for the mixed-Higgs scenario (the heavy CP-even Higgs is above the LEP
  limit on the mass of the SM Higgs boson and the coupling squared of
  the light Higgs to $ZZ$ is below the LEP limit, $\xi^2 \lesssim
  0.25$ for $m_h \sim 100$ GeV).  }
\label{fig:tb_10}
\end{figure}

For given $\tan \beta$ and $m_A$ we can determine the minimal value of
the SUSY correction needed for both scenarios to be viable.  In
Fig.~\ref{fig:tb_10}, we plot contours of constant $m_h$, $m_H$ and
$\xi^2$ in the $m_A$ -- $r_\Delta$ plane for $\tan \beta = 10$.  We
easily recognize the behavior of Higgs masses. In the decoupling
limit, $m_A \gg m_Z$, the mass of the light CP-even Higgs boson does
not depend on $m_A$ and the mass of the heavy CP-even Higgs boson scales
linearly with $m_A$. The masses of both Higgses increase with increased
radiative correction to the 2-2 element of the Higgs mass-squared
matrix.  We see that the mixed-Higgs scenario is viable for $r_\Delta
\gtrsim 0.5$ while the decoupled scenario requires $r_\Delta \gtrsim
0.65$. Thus, the mixed-Higgs scenario requires a somewhat smaller
contribution from the stop sector.  On the other hand, this scenario
works only in a limited range of $m_A$ since the soft-SUSY-breaking
parameters $\mhdsq$ and $\mhusq$ have to be
carefully adjusted relative to one another given the relation
\begin{equation}
m_A^2 \simeq m_{H_d}^2 - m_{H_u}^2 - m_Z^2
\label{marel}
\end{equation}
at  tree level for $\tan \beta > {\rm few}$. In the above equation,
$\mhdsq$ and $\mhusq$ are the weak-scale values. 
The particular values to which they  evolve
at the GUT scale will depend on many of the other weak-scale parameters.  
Thus, in order to achieve the required
relation of Eq.~(\ref{marel}), all of the various GUT-scale parameters
will have to be closely correlated in a very particular way. Furthermore,
as we will see, the improvement in naturalness is limited to a very
small window in $\tan \beta$, which implies also a small window for the
 $B_\mu$ parameter due to the relation
\begin{equation}
m_A^2 \simeq B_\mu \tan \beta.
\end{equation}

\begin{figure}
\includegraphics[width=2.5in]{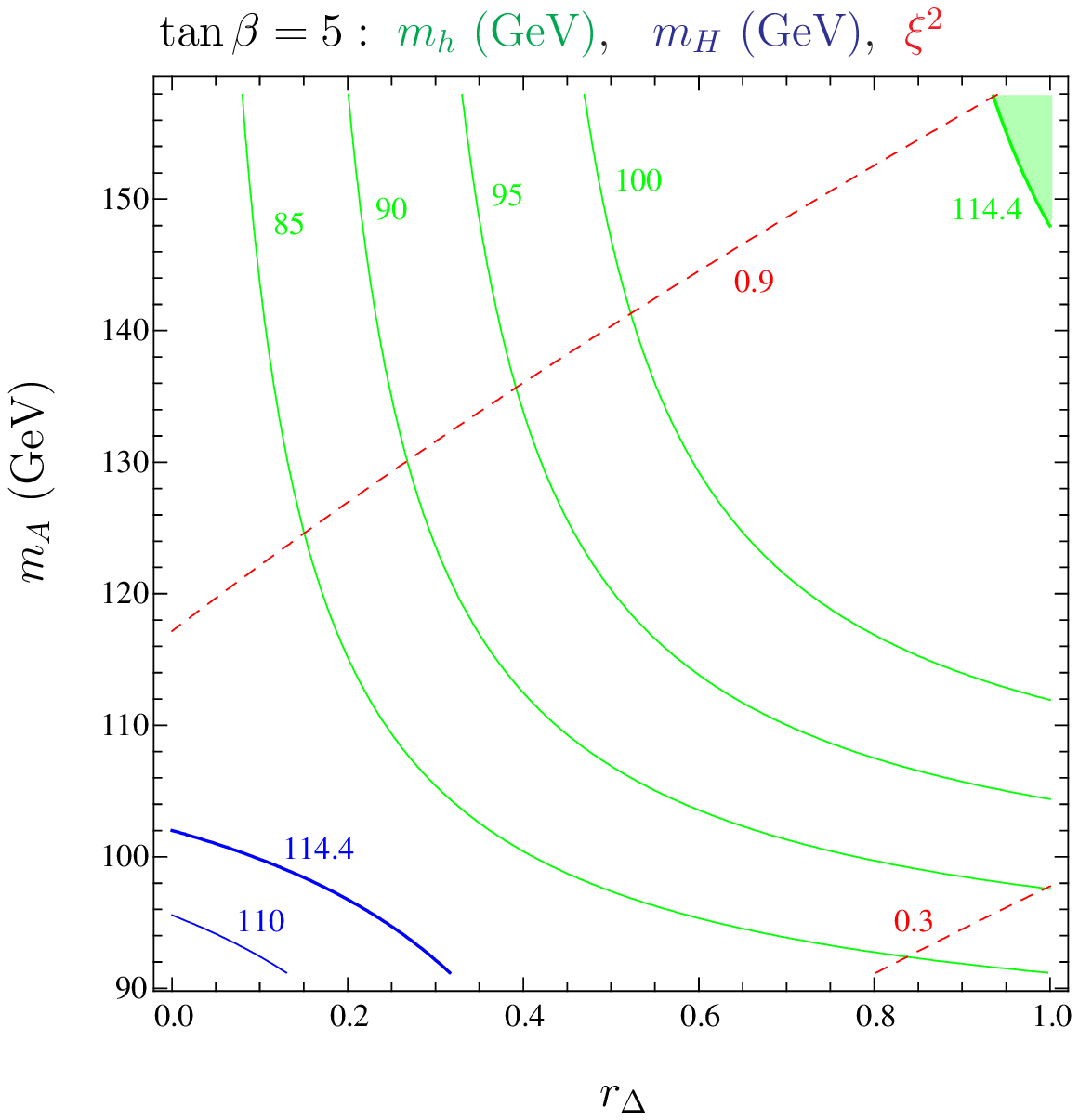}
\includegraphics[width=2.5in]{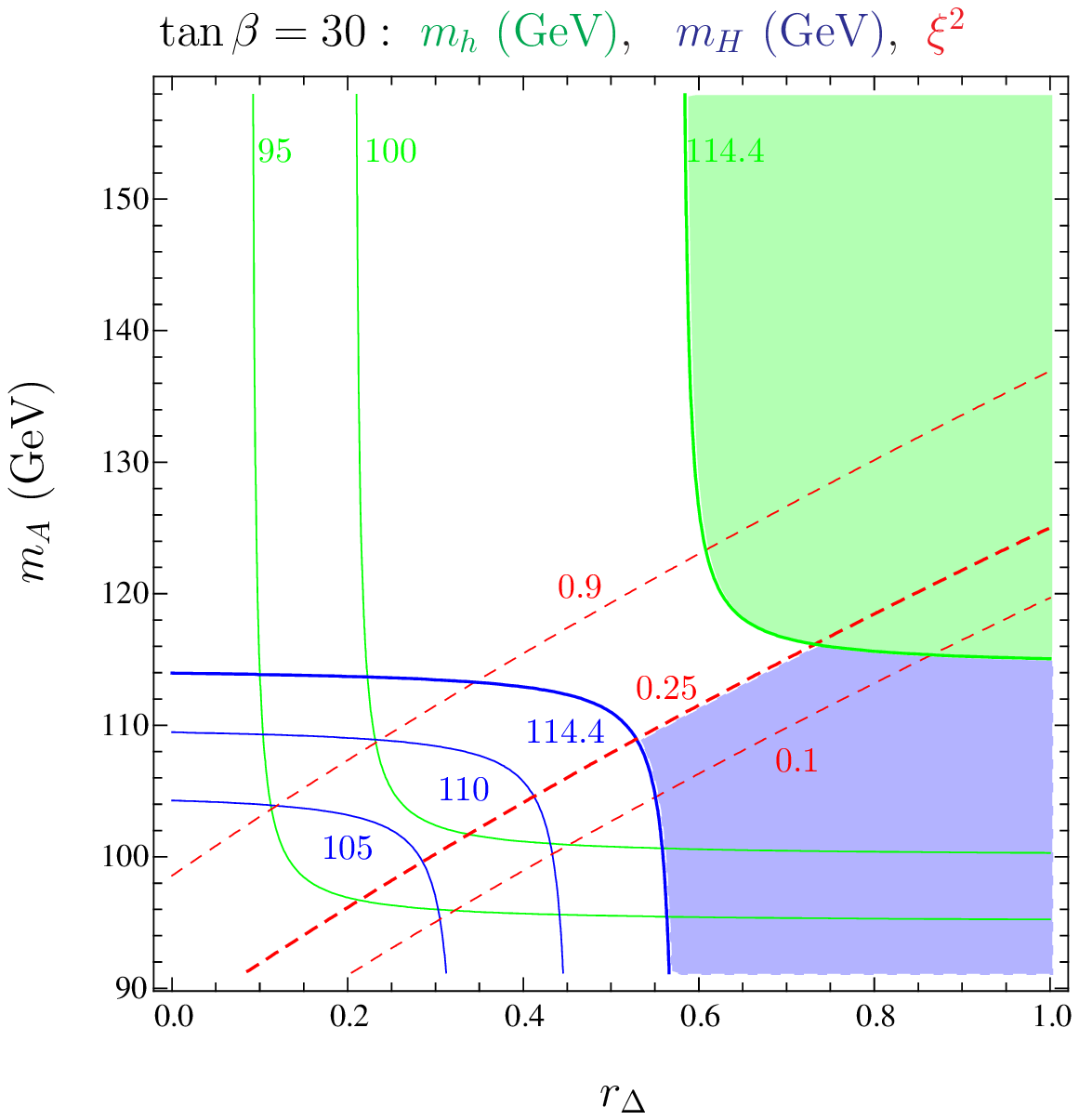}
\caption{
  Contours of constant $m_h$ (green), $m_H$ (blue) and $\xi^2$ (red)
  in $m_A$ -- $r_\Delta$ plane for $\tan \beta = 5$ (top) and $\tan
  \beta = 30$ (bottom). The meaning of the shaded regions is the same
  as in Fig.~\ref{fig:tb_10}.  }
\label{fig:tb_5and30}
\end{figure}

For smaller $\tan \beta$, the contribution from mixing is more
significant and so the heavy Higgs can be heavier than 114.4 GeV for a
smaller value of the SUSY correction, $\Delta$. However, the
off-diagonal term in $M$ is increased and Higgs mixing is larger,
which significantly increases $\xi^2$. In Fig.~\ref{fig:tb_5and30}
(top) we see that the mixed-Higgs scenario is not viable for $\tan
\beta = 5$. (Note that we do not show $\mha<90\gev$ in the plots. This
is because at fixed $r_\Delta$ both $\mhl$ and $\mhh$ decrease as
$\mha$ decreases and $Z\to \hl+\ha$ limits from LEP enter.)  For $\tan
\beta$ significantly above $10$, the Higgs-mixing induced by the
off-diagonal element of $M$ is negligible and thus the mixed-Higgs
scenario requires basically the same radiative correction as the
decoupled one.  This is clearly visible in the behavior of the masses
of the light and heavy CP-even Higgses for $\tan \beta = 30$, see
Fig.~\ref{fig:tb_5and30} (bottom).

\begin{figure}
\includegraphics[width=2.5in]{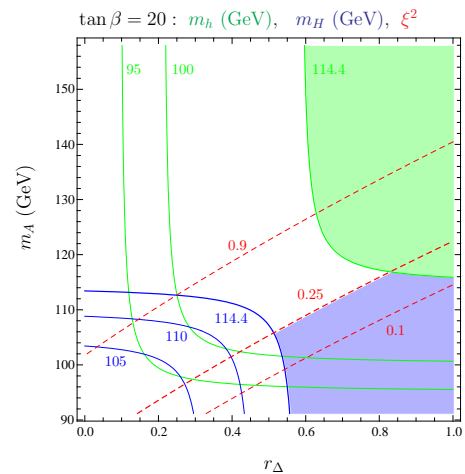}
\caption{
  Contours of constant $m_h$ (green), $m_H$ (blue) and $\xi^2$ (red)
  in $m_A$ -- $r_\Delta$ plane for $\tan \beta = 20$. The meaning of
  shaded regions is the same as in Fig.~\ref{fig:tb_10}.  }
\label{fig:tb_20}
\end{figure}

It is worth noting that for $\tanb\sim 20$ the mixed-Higgs scenario
can yield both the 98 GeV and the 116 GeV excesses of Higgs-like
events observed at LEP. This is illustrated in Fig.~\ref{fig:tb_20}.
For $\tan \beta = 20$, we see that in the vicinity of
$[\mha,r_\Delta]\sim [100\gev,0.6]$ the mass of the light Higgs is
about 98 GeV with $\xi^2 \sim 0.1$ (as needed to explain the excess of
Higgs-like events at 98 GeV) while the heavy Higgs has a mass of about
116 GeV (and $g_{ZZH}^2/g_{ZZ\hsm}^2\sim 0.9$). This possibility was
studied in detail in Ref.~\cite{Drees:2005jg}. It is clear from
Fig.~\ref{fig:tb_20} that this mixed-Higgs scenario that explains
simultaneously both the 98 GeV and 116 GeV LEP 
excesses of Higgs-like events requires basically the same size
of SUSY correction, $\Delta$, as the decoupled scenario and thus it works in the
same region of SUSY parameter space.

\begin{figure}
\includegraphics[angle=90,width=3.in]{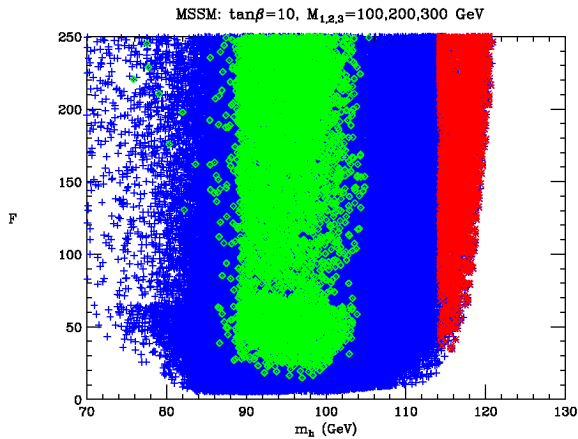}
\caption{
  Fine tuning vs. $m_h$ for randomly generated MSSM parameter choices
  with $\tan \beta=10$ and $M_{1,2,3}(m_Z)=100,200,300\gev$.  Blue pluses
  correspond to parameter choices yielding $m_h<114$ GeV that are
  ruled out by LEP limits on the Higgs mass and as a function of 
the $ZZh$ coupling.
Green diamonds are the mixed-Higgs scenarios with $m_h<114\gev$
  GeV that satisfy LEP limits due to reduced $ZZh$ coupling. 
Red crosses are points with $m_h>114$ GeV --- these 
  automatically satisfy LEP limits. }
\label{fig:fvsmh}
\end{figure}

\begin{figure}
\includegraphics[angle=90,width=3.in]{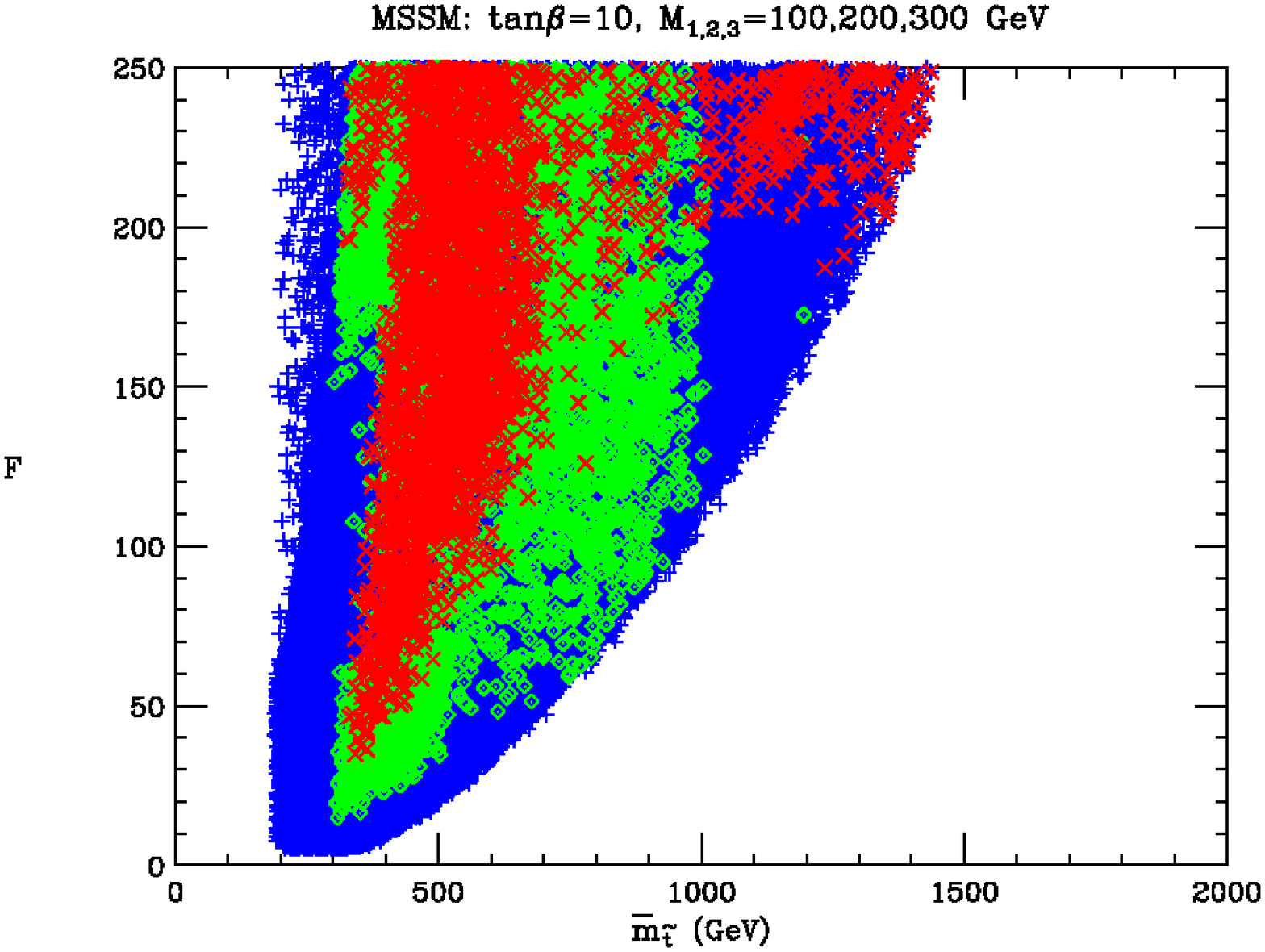}
\includegraphics[angle=90,width=3.in]{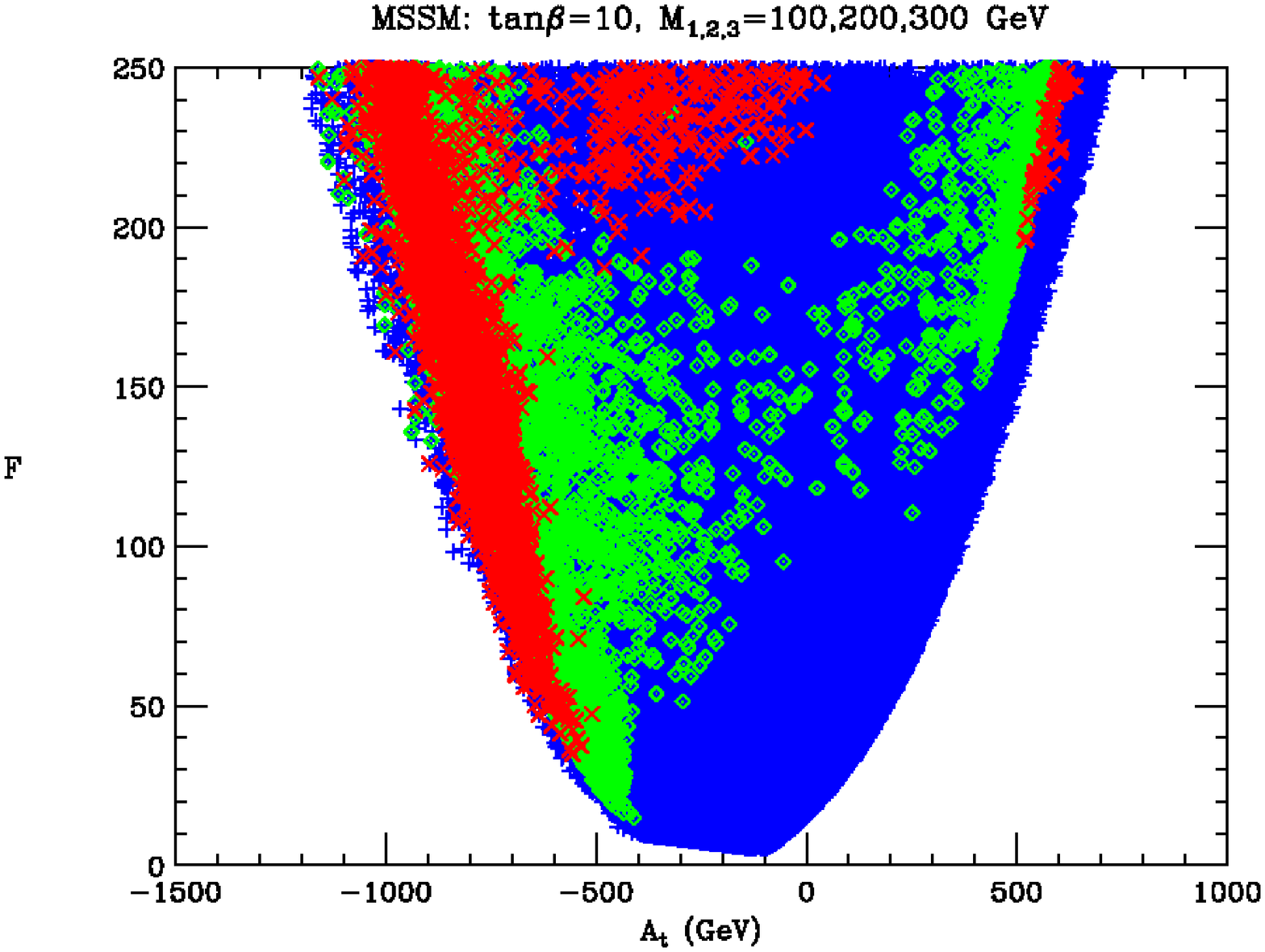}
\caption{
Fine tuning vs. ${\bar m}_{\tilde t}$ (top) and fine tuning vs. $A_t$ (bottom) for randomly generated MSSM parameter choices
  with $\tan \beta=10$ and  $M_{1,2,3}(m_Z)=100,200,300\gev$. Point convention as in Fig.~\ref{fig:fvsmh}.
}
\label{fig:mssm_mstop_and_At}
\end{figure}

From this simplified exercise we thus learn that LEP consistency of the
mixed-Higgs scenario still requires a significant correction from the
stop sector although somewhat smaller than the decoupled scenario.
Now we proceed with a precise numerical analysis of the associated
fine tuning which closely follows
the analysis outlined in Ref.~\cite{Dermisek:2007yt}.  Compared to our
previous work, we designed a special scan to pick up mixed-Higgs
scenarios which would occur very rarely in a random scan due to the
relatively narrow range of $m_A$, or, alternatively, of $m_{H_d}$,
required.  In these scans, we employ the fixed value of $\tan \beta =
10$ (which our discussion has shown should give the most improvement
on fine tuning relative to the decoupled scenario) and fixed gaugino
soft masses of $M_{1,2,3}=100,200,300\gev$. We scan over all other
soft-SUSY-breaking parameters, including $\mu$, $B_\mu$
and the third generation (stop)
parameters $m_Q$, $m_U$, $m_D$, and $A_t$, all
defined at scale $\mz$.  For each set of these $\mz$-scale parameter
choices, we determine the values of all the soft parameters at the the
GUT scale, $\mgut\sim 2\times 10^{16}\gev$, by renormalization group
evolution. We then vary each GUT scale parameter, $p$, in turn, and
evolve back to scale $\mz$ to determine how much $\mz$ has changed.
From this we compute $F$ of Eq.~(\ref{fdef}). The resulting values of $F$ are presented in
Figs.~\ref{fig:fvsmh} and~\ref{fig:mssm_mstop_and_At} as a function of
$\mh$, $\mstopbar$ and $A_t$.  A blue $+$ is plotted whenever there is
a soft-SUSY-breaking scenario with $\mh<114\gev$ that is excluded by
LEP due to the fact that $\xi^2$ is too large.  Overlaid on the blue
$+$'s we plot a green diamond whenever there is a choice of
soft-SUSY-breaking parameters yielding $\mh<114\gev$ but with
sufficiently reduced $ZZh$ coupling (due to the effects of Higgs
mixing) so as to not be excluded by LEP . A red $\times$ is plotted
whenever there is a soft-SUSY-breaking parameter set yielding
$\mh>114\gev$ --- LEP constraints are automatically satisfied in this
case.

In Fig.~\ref{fig:fvsmh} we see that the mixed-Higgs scenarios (green
diamonds) require $m_A$ to be near $\sim 90 - 100$ GeV; fine tuning
can be as low as $F\sim 15.5$ (6.5\% parameter tuning), 
a significant reduction compared to the
decoupled scenario (red crosses) for which the minimal $F$ is about
30. The least fine-tuned decoupled scenarios require large mixing in
the stop sector, as shown in Fig.~\ref{fig:mssm_mstop_and_At}. From
the same figure, we see that the mixed-Higgs scenarios extend the
region of SUSY parameter space allowed by the LEP constraints to
slightly smaller stop masses and somewhat smaller mixing.

In conclusion, the level of fine tuning in the mixed-Higgs scenario can be
reduced to about 6.5\% compared to the 3\% fine tuning needed in the
decoupled Higgs scenario. However, this improvement happens only in a
limited range of $\tan \beta$ (the results presented for $\tanb =
10$ are close to the optimal choice) and $m_A$. As a result,
 the mixed-Higgs scenario
requires additional constraints on the $m_{H_d}$ and $B_\mu$ parameters
which are not constrained in the case of the decoupled Higgs scenario. For
smaller $\tan \beta$ the mixed-Higgs scenario is not viable and for
large $\tan \beta$ it requires the same level of fine tuning as the
decoupled Higgs solution.

The fact that the required magnitude of $\Delta$ is similar in the
mixed and decoupled scenarios means that they both prefer the same
region of SUSY parameter space, namely that with large mixing in the
stop sector. The mixed-Higgs scenario allows for continuation of this
region to somewhat smaller mixing for optimal $\tan \beta$. The
large mixing in the stop sector can be achieved either in models which
generate a large top soft trilinear coupling, $A_t$, at a large scale or
it can be achieved by renormalization group evolution in models which
generate negative stop masses squared at a large
scale~\cite{Dermisek:2006ey}.

\section{Mixed and Unmixed Higgs Scenarios in the NMSSM}
\label{sec:singletcases}

The NMSSM is an extremely attractive model \cite{ourfirstpaper,allg}.
In particular, it provides a very elegant solution to the $\mu$ problem of the
MSSM via the introduction of a singlet superfield $\widehat{S}$.  For
the simplest possible scale invariant form of the superpotential, the
scalar component of $\widehat{S}$ naturally acquires a vacuum
expectation value of the order of the \susy\ breaking scale, giving
rise to a value of $\mu$ of order the electroweak scale. The NMSSM is
the simplest supersymmetric extension of the standard model in which
the electroweak scale originates from the \susy\ breaking scale only.

Apart from the usual quark and lepton Yukawa couplings, the scale
invariant superpotential of the NMSSM is
%\vspace*{-.07in}
%\beq \label{1.1}
$
W=\lambda \widehat{S} \widehat{H}_u \widehat{H}_d + \third\kappa
\widehat{S}^3
$ 
%\vspace*{-.11in}
%\eeq 
depending on two dimensionless couplings $\lambda$, $\kappa$ beyond
the MSSM.  [Hatted (unhatted) capital letters denote superfields
(scalar superfield components).]  The associated trilinear soft terms
are
%
%\vspace*{-.1in}
%\beq \label{1.2}
$
\lambda A_{\lambda} S H_u H_d + \third\kappa A_\kappa S^3 \,. 
$
%\vspace*{-.1in}
%\eeq
%
The final two input parameters are 
%
%\vspace*{-.1in}
%\beq \label{1.3} 
$
\tan \beta = h_u/h_d$ and $\mu_\mathrm{eff} = \lambda
s \,, 
$
%\vspace*{-.07in}
%\eeq 
%
where $h_u\equiv
\vev {H_u}$, $h_d\equiv \vev{H_d}$ and $s\equiv \vev S$.
The Higgs sector of the NMSSM is thus described by the six parameters
%\vspace*{-.1in}
%\beq \label{6param}
$\lambda\ , \ \kappa\ , \ A_{\lambda} \ , \ A_{\kappa}, \ \tan \beta\ ,
\ \mu_\mathrm{eff}\ .$
%\vspace*{-.1in}
%  \eeq
In addition, values must be input for the gaugino masses 
and for the soft terms related to the (third generation)
squarks and sleptons that contribute to the
radiative corrections in the Higgs sector and to the Higgs decay
widths. 

The particle content of the NMSSM differs from
the MSSM by the addition of one CP-even and one CP-odd state in the
neutral Higgs sector (assuming CP conservation), and one additional
neutralino.  The result is three CP-even Higgs bosons ($h_{1,2,3}$)
two CP-odd Higgs bosons ($a_{1,2}$) and a total of
five neutralinos $\wtil\chi^0_{1,2,3,4,5}$. While we
denoted the CP-even and CP-odd neutral Higgs bosons of the MSSM as $h,H$
and $A$, respectively, those of the NMSSM will be denoted by
$\hi,\hii,\hii$ and $\ai,\aii$, respectively. In the latter case, our
focus will be on the lightest states $\hi$ and $\ai$. 
 The NMHDECAY program \cite{Ellwanger:2004xm}, which
includes most LEP constraints, allows easy exploration
of Higgs phenomenology in the NMSSM.

The NMSSM study presented in this paper focuses on cases in which the
lightest Higgs boson can have $\mhi<114\gev$ without violating LEP
limits and without necessarily having the $\hi\to \ai\ai$, with
$\mai<2\mb$ decay being dominant.  While it is true that this latter
situation gives rise to models with the very least fine tuning, there
are alternative models with only modest fine tuning in which
$\mhi<114\gev$ but substantial Higgs mixing suppresses the $ZZ\hi$
coupling sufficiently that the $\epem\to Z^*\to Z\hi$ production rate
is reduced to an allowed level even if $\hi\to b\anti b$ and/or
$\hi\to\ai\ai\to 4b$ decays are dominant.  

This can occur in a number of ways. The first possibility is that the
$\hi$ has substantial singlet $S$ component.  In such scenarios, it is
typically the $\hii$ that is the most SM-like CP-even Higgs boson, but
$\mhii>114\gev$ and LEP constraints do not apply to the $\hii$.
Another possibility is the analogue of the MSSM mixed-Higgs scenarios
described in the preceding MSSM sections. For these, the $\hi$ and
$\hii$ both have mass near $100\gev$ and are primarily non-singlet,
but mix in such a way that the LEP limits are evaded. And, of course,
there are LEP-allowed scenarios in which the $\hi$ mixes partly with
the singlet and partly with the other MSSM-like Higgs boson.  We have
performed a broad scan over NMSSM parameter space to look for and
investigate the fine tuning associated with scenarios of each type.
As discussed below, not all the points of this type found in our scans
are highly fine-tuned.  There are specific parameter regions that
produce points of each type that are only moderately fine-tuned for
which the $\hi$ has $\mhi<114\gev$ but escapes LEP limits by virtue of
small $ZZ\hi$ coupling.

\begin{figure}[ht!]

  \centerline{\includegraphics[width=2.4in,angle=90]{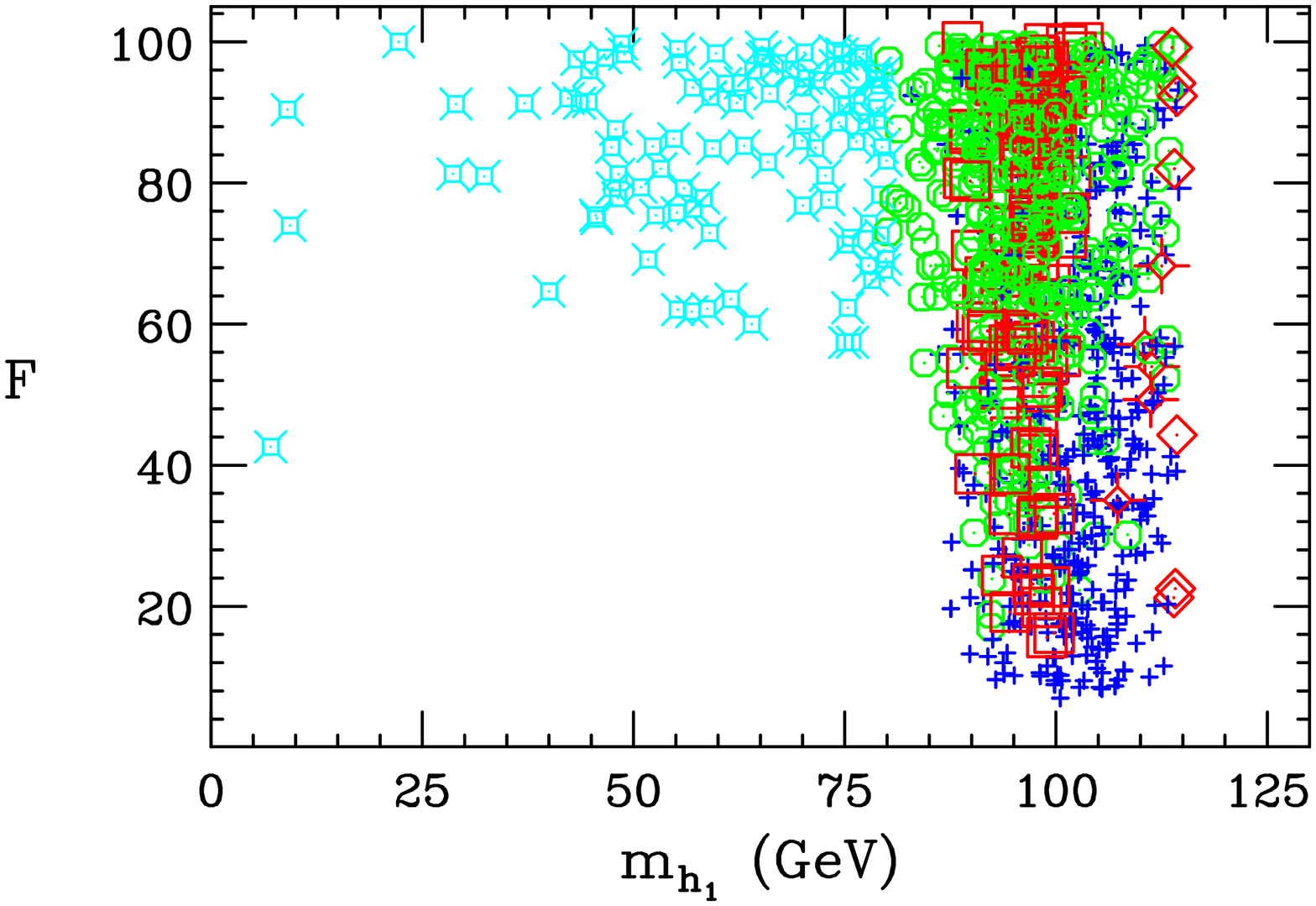}}
  \vspace*{.1in}
  \centerline{\includegraphics[width=2.4in,angle=90]{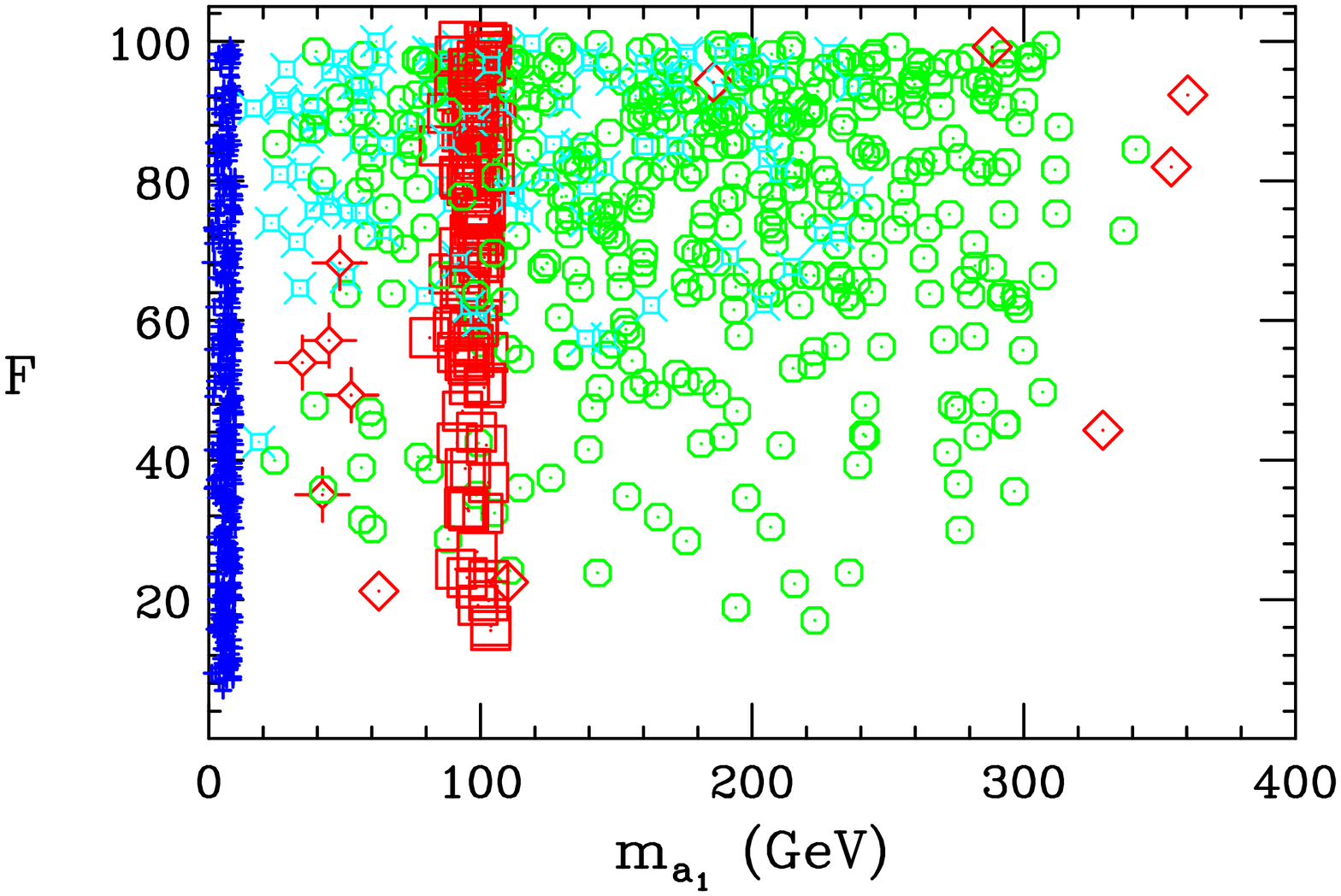}}
  \vspace*{.1in}
  \centerline{\includegraphics[width=2.4in,angle=90]{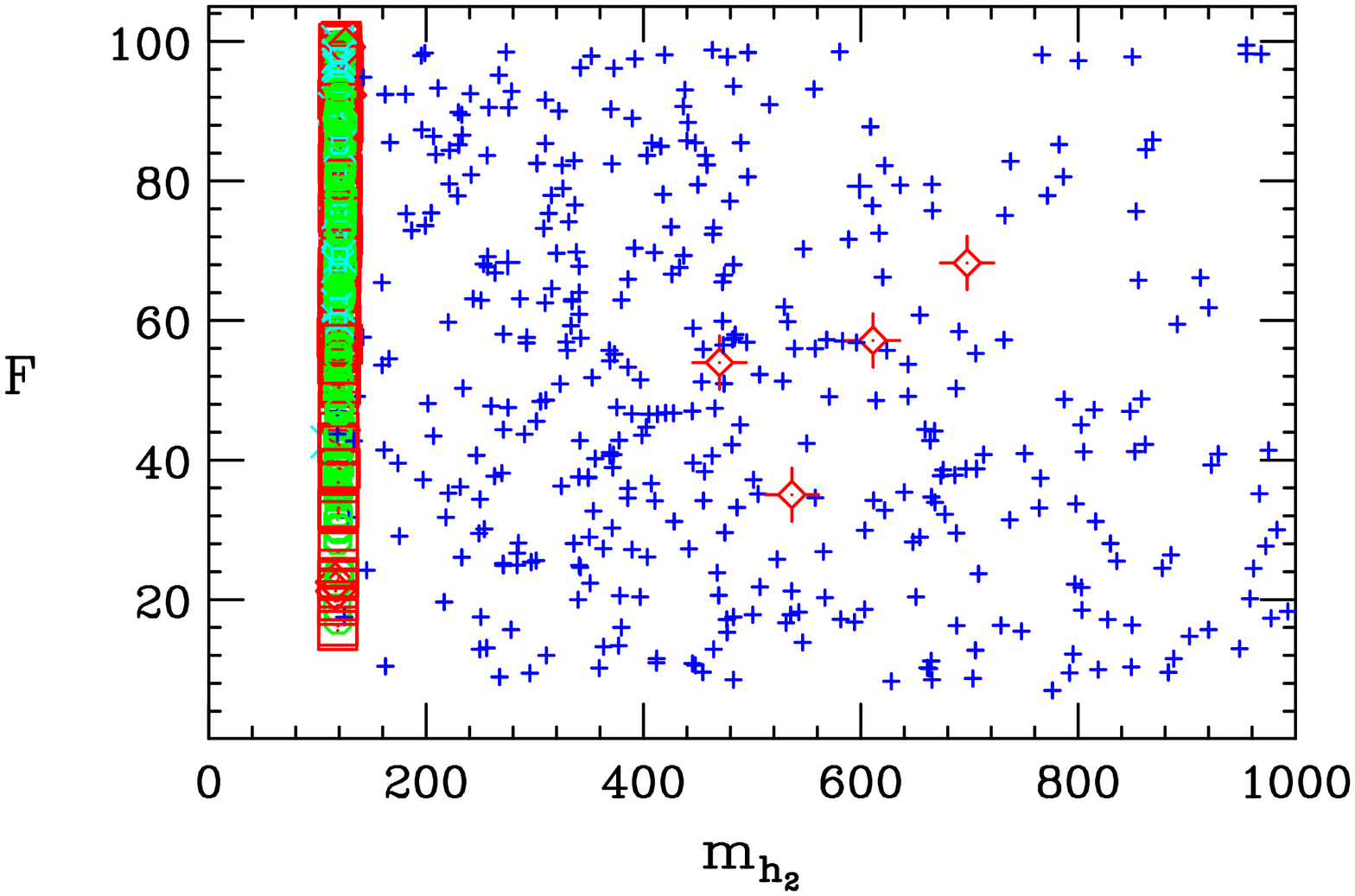}}
\caption{ For fixed $M_{1,2,3}(\mz)=100,200,300\gev$
and $\tanb=10$ we plot: $F$ vs. $\mhi$ (top); $F$ vs. $\mai$
(middle); and $F$ vs. $\mhii$ (bottom).
In this and all succeeding plots, all points have $F<100$ and
$\mhi<114\gev$.  The blue $+$ points are ones with a very SM-like
$ZZ\hi$ coupling that escape LEP limits because
$\mai<2\mb$  and $\hi\to\ai\ai\to 4\tau$ or $4j$ decays are
dominant.  All other points have $\mai>2\mb$.  The definitions of 
the other points appear in the text.}
\label{mixplots1}
\vspace*{-.1in}
\end{figure}

\begin{figure}[ht!]
  \centerline{\includegraphics[width=2.4in,angle=90]{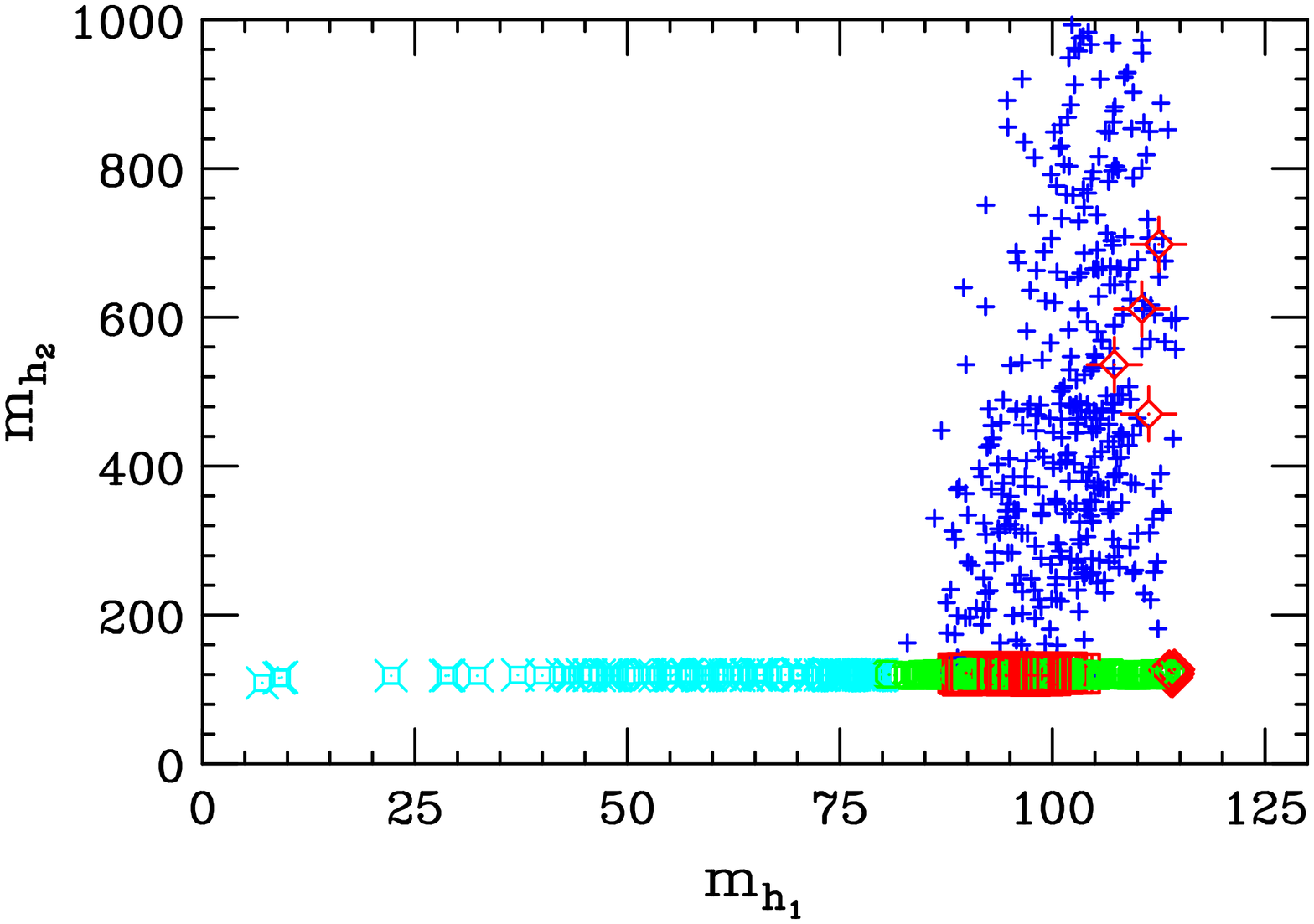}}
  \centerline{\includegraphics[width=2.4in,angle=90]{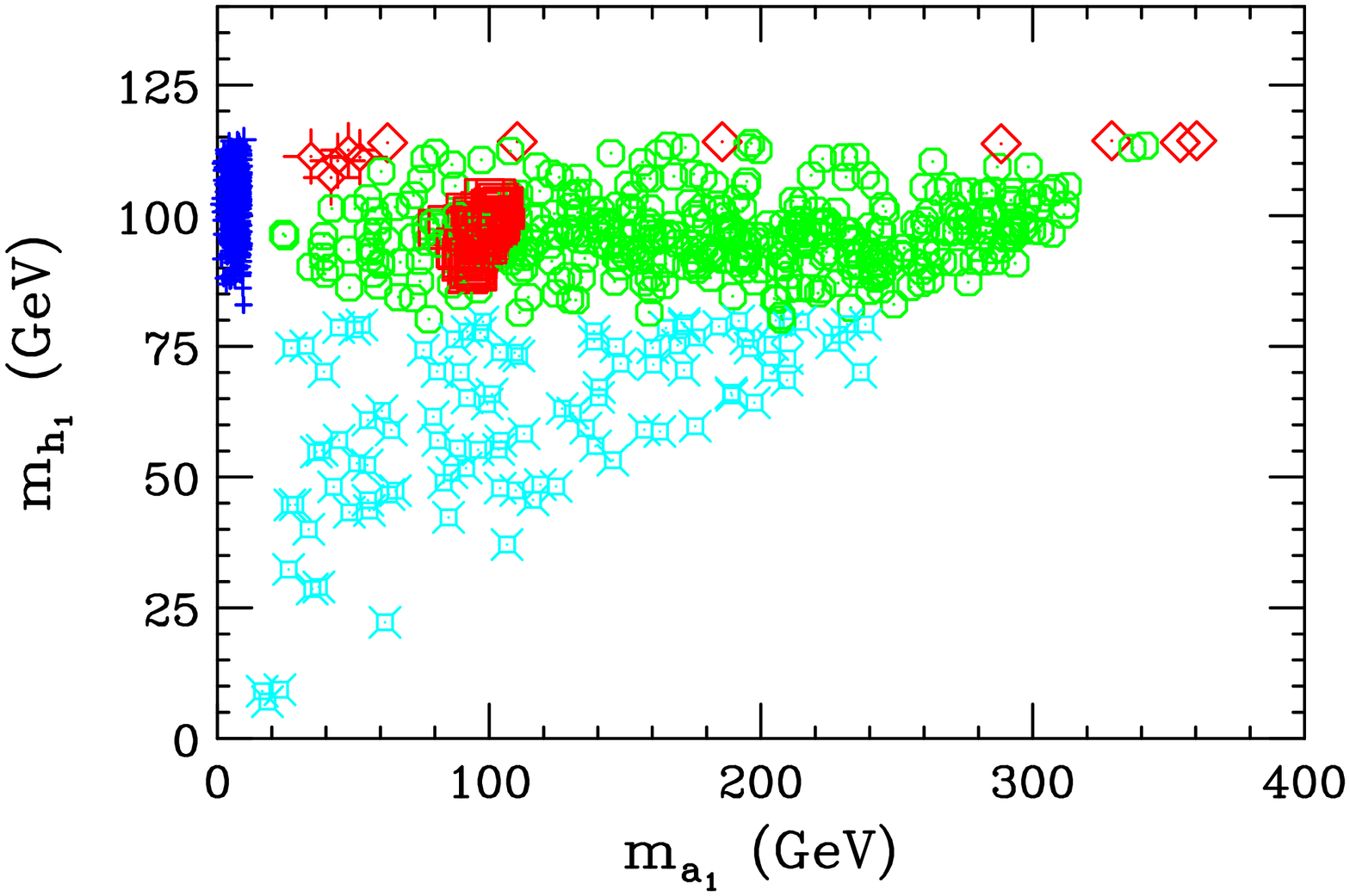}}
\caption{ For fixed $M_{1,2,3}(\mz)=100,200,300\gev$
and $\tanb=10$ we plot: $\mhii$ vs. $\mhi$ (top) and $\mhi$ vs. $\mai$
(bottom). Notation and conventions as in Fig.~\ref{mixplots1}.
}
\label{mhimhii}
\vspace*{-.1in}
\end{figure}

\begin{figure}[ht!]
  \centerline{\includegraphics[width=2.4in,angle=90]{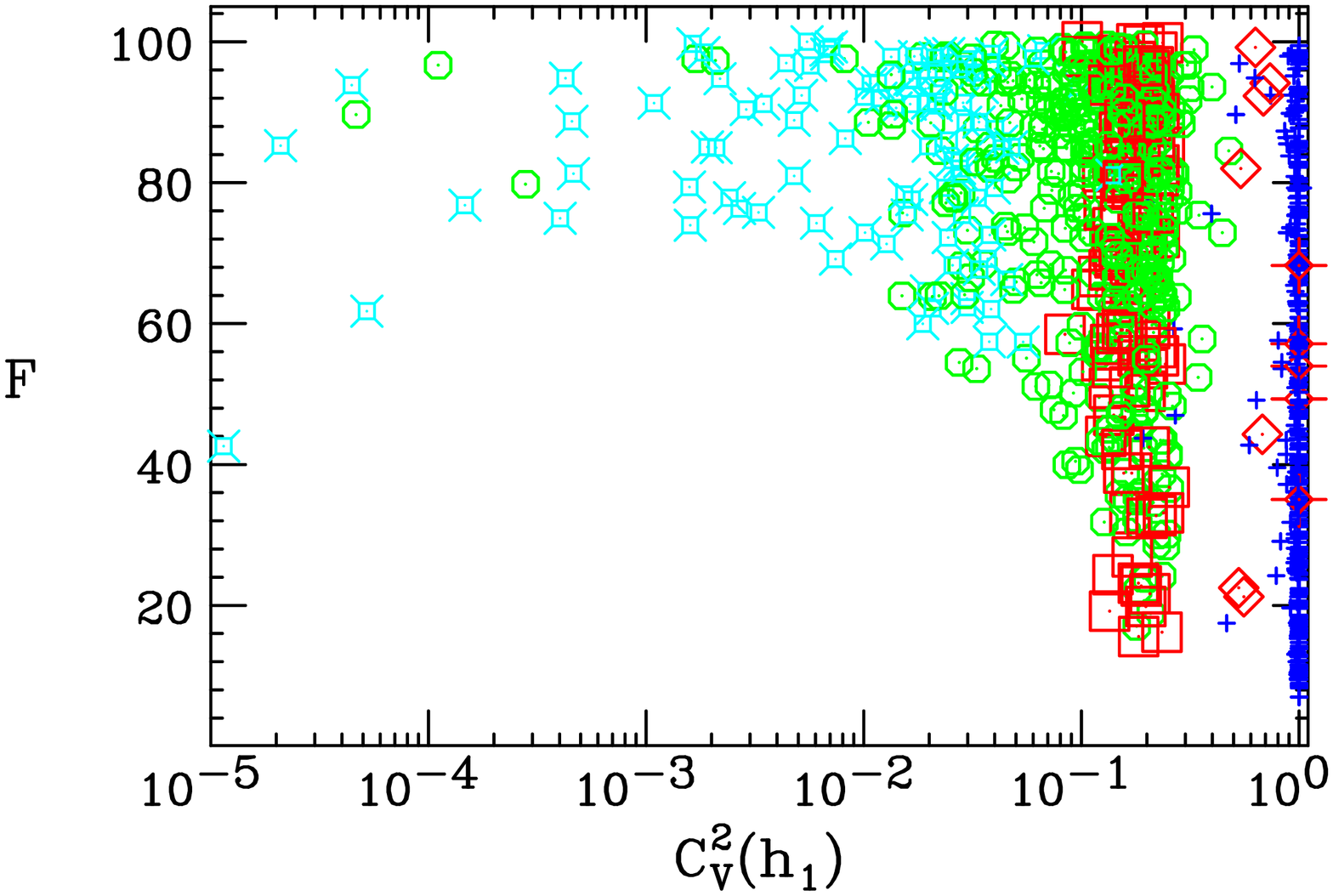}}
  \centerline{\includegraphics[width=2.4in,angle=90]{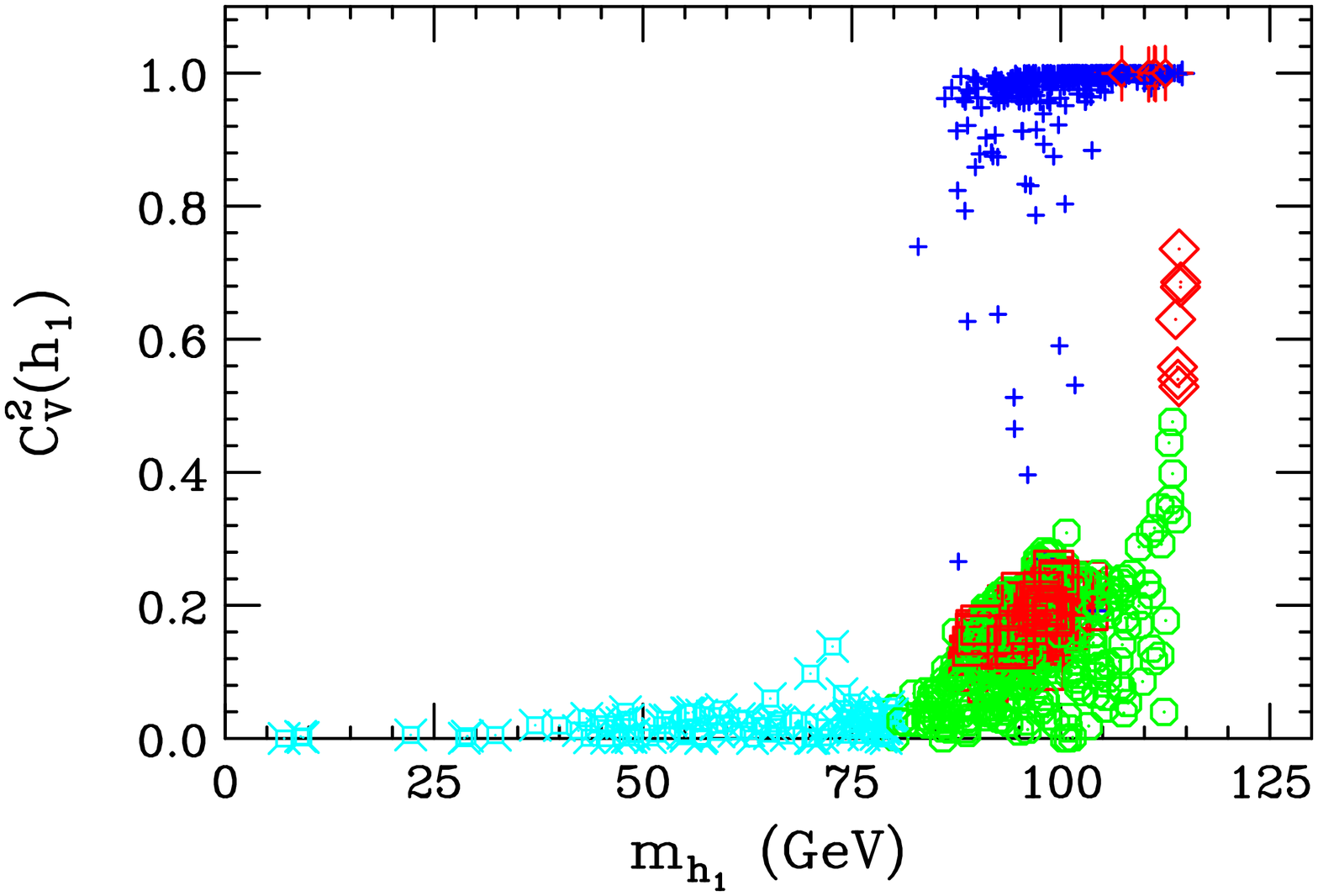}}
  \centerline{\includegraphics[width=2.4in,angle=90]{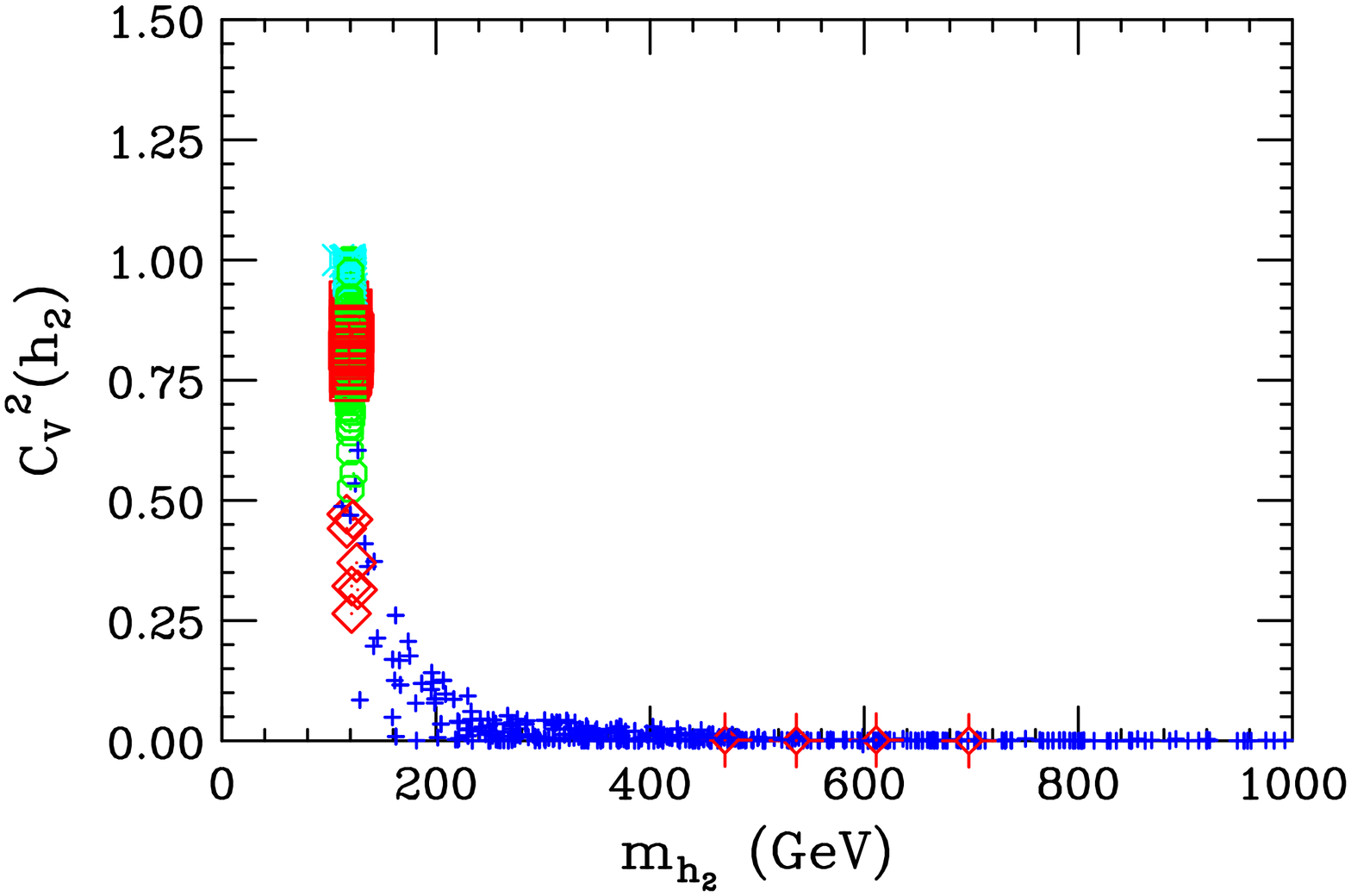}}
\caption{For fixed $M_{1,2,3}(\mz)=100,200,300\gev$
and $\tanb=10$ we plot $F$ vs. $C_V^2(\hi)$, $C_V^2(\hi)$ vs. $\mhi$
and $C_V^2(\hii)$ vs. $\mhii$.  Notation and conventions as in Fig.~\ref{mixplots1}.
}
\label{mixplots2}
\vspace*{-.1in}
\end{figure}

To be explicit, let us take $M_{1,2,3}=100,200,300\gev$ and
$\tanb=10$, as before. We scan over a broad range of all other NMSSM
parameters searching for points that: a) are consistent with all
constraints built into NMHDECAY; b) obey the additional requirement
that the effective $Z+b's$ rate from $Z\hi$ production, as quantified
via 
\bea 
\xi^2(Z+b's)&\equiv& {g_{ZZ\hi}^2\over
  g_{ZZ\hsm}^2}\Bigl[\br(\hi\to b \anti b)\cr
&&+\br(\hi\to\ai\ai)\left[\br(\ai\to b\anti b)\right]^2\Bigr]\,, 
\eea
lies below the LEP limit on the $Z+2b$ final state.\footnote{This
  constraint can be stronger than necessary in cases where the $4b$
  final state is dominant, for which it is known that LEP limits only
  imply $\mhi\lsim 110\gev$.  However, absent full LEP analysis on a
  point-by-point basis it is the safest approach available to us.
  Additional allowed points might emerge in a point-by-point
  approach.}  Fig.~\ref{mixplots1} shows the electroweak fine-tuning
measure $F$ as a function of $\mhi$, $\mai$ and $\mhii$.  In this, and
all succeeding plots, we only show points with $F<100$, corresponding
to fine tuning no worse than $1\%$.  One sees (the blue $+$'s) the
expected large number of points with low $F$, $\mhi\sim 100\gev$ and
$\mai<2\mb$ that escape LEP limits by virtue of large
$\br(\hi\to\ai\ai\to 4\tau)$.~\footnote{The very broad scans focused
  on mixed-Higgs scenarios performed for this paper did not pick up
  the very lowest $F$ points that have $F\sim 6$ found in our
  specialized scans of earlier papers.}  In addition, there are
several classes of points with only somewhat higher minimum $F$ that
escape LEP limits. We detail these below. We note that the density of
points in the various classes we shall discuss is somewhat a function
of how we did the scanning.  For instance, we worked hard to find
MSSM-like mixed-Higgs scenarios, whereas we did not do so for the
other mixed-Higgs scenarios.  And some scan runs purposely
deemphasized the (blue) $+$ points that previous papers have focused
on.

First, there are the (red) diamond-star points with an essentially
pure singlet $\ai$ with $\mai\sim 50\gev$ and an $\hi$ with very
SM-like $ZZ\hi$ coupling and $\mhi\sim 110\gev$ that escape LEP
published limits by virtue of $\hi\to \ai\ai\to 4\gam$ being the
dominant $\hi$ decay. As discussed in our previous paper, the
fine tuning of $\alam$ and $\akap$ 
needed to achieve an almost purely singlet $\ai$ is quite
significant and, further, it is likely that a LEP analysis of the
$Z+4\gam$ final state would eliminate these points.  Nonetheless, we
include them since they can have $F$ as small as about $35$, corresponding to
about $3\%$ parameter tuning to get proper EWSB.

The remaining points have $\mai>2\mb$ (and $\mhi<114\gev$) 
and escape LEP limits
by virtue Higgs-mixing leading to suppressed $ZZ\hi$ coupling. 
A discussion of the $ZZ\hi$ coupling is appropriate before giving our
classification of these points. 
Defining
\beq
C_V^2(\hi)\equiv {g_{ZZ\hi}^2\over g_{ZZ\hsm}^2}\,,
\eeq
one has
\beq
\cvisq=(\sinb S_{11}+\cosb S_{12})^2\,,
\eeq
where the $\hi$ mixture is defined by
\bea
\hi=S_{11}H_{u\,R}+S_{12}H_{d\,R}+S_{13} S_R\,,
\eea
and similarly for $\hii$ and $\hiii$.  Here,
the neutral Higgs fields are taken to be
\bea
H_u^0&=&h_u+{H_{u\,R}+iH_{u\,I}\over \sqrt 2}\nn\\
H_d^0&=&h_d+{H_{d\,R}+iH_{d\,I}\over \sqrt 2}\nn\\
S&=&s+{S_R+iS_I\over \sqrt 2}\,,
\eea
with $h_u,h_d,s$ being the vevs.
We will similarly write
\beq
\ai=P_{11}\left(\cosb H_{u\,I}+\sinb H_{d\,I}\right)+P_{12} S_I\,,
\eeq
and similarly for $\aii$.
When $\tanb$ is large (as it is for this $\tanb=10$ discussion),
$\cosb$ is small and if $S_{11}$ is small then $\cvisq\ll 1$ is
automatic.

In the figures, we have divided the remaining scenarios into four
distinct categories.  
\ben
\item The first large group of points (indicated by large cyan starred
  squares) are those for which $\mhi<80\gev$ (including very small
  $\mhi$) and the $\hi$ is largely singlet, $|S_{13}|\sim 1$. The
  $\hii$ has $C_V^2(\hii)\sim 1$ but escapes LEP limits since
  $\mhii>114\gev$; in fact, almost invariably $\mhii\sim 120\gev$ for
  these points, with a few having $\mhii$ between $110\gev$ and
  $118\gev$. The minimum $F$ for this category of point is $F\sim 40$.
\item The second large group of points (indicated by large green, or
  darker, circles) have $\mhi>80\gev$ (clustered about $\mhi\sim
  100\gev$) but have sufficiently suppressed $C_V^2(\hi)$ because the
  $\hi$ is predominantly singlet: $(S_{11}^2+S_{12}^2)<0.5$.  These
  points have $|P_{11}|\sim 0$, implying that the $\ai$ is very nearly
  pure singlet. The $\hii$ has $\mhii\sim 120\gev$ and
  $C_V^2(\hii)>0.5$ for these points. The minimum $F$ for these points
  is $F\sim 17$, or $6\%$ fine tuning, which, while not as good as the
  blue $+$ points (which can reach down to $F\sim 6$ in a fuller
  scan), is not really too
  bad.
\item The third set of points are the large (red) plain diamonds.  For
  these points, $0.5\leq (S_{11}^2+S_{12}^2)<0.9$, $|S_{12}|<0.1$,
  implying that the $\hi$ is mainly $H_{u\,R}$, and $\mhi$ is just
  below $114\gev$.  Almost any value of $\mai\gsim 60\gev$ (implying
  no $\hi\to \ai\ai$ decays) is possible and the $\ai$ is nearly
  purely singlet. The minimum $F$ here is $F\sim 22$. These scenarios
  are remnants of the usual decoupled scenarios (for which
  $\mhi>114\gev$ and for which we found $F\sim 20$ in
  Ref.~\cite{Dermisek:2007yt}) --- they instead have Higgs mass just
  slightly below $114\gev$ and just enough Higgs mixing to escape LEP
  limits.

\item Points in the fourth and final set are those that are the NMSSM
  analogues of the MSSM points with strong mixing between the two
  doublet Higgs fields. These are indicated by the large (red) plain
  squares. These have $\mai\sim\mhi\sim 100\gev$,
  $(S_{11}^2+S_{12}^2)>0.9$ and $|P_{11}|\sim 1$, implying that both
  the $\ai$ and the $\hi$ have small singlet component. Typically,
  $|S_{12}| \sim 0.95$ and $|S_{11}|\sim 0.3$, implying that the $\hi$
  is mainly $H_{d\,R}$. The minimum $F$ for these MSSM-like
  mixed-Higgs scenarios found in our scans is $F\sim 16$. This is the
  same level as achieved for the mixed-Higgs scenarios in the MSSM
  scans discussed previously.

\een
With regard to the 4th category above, it is useful to recall from
Ref.~\cite{ourfirstpaper} that the MSSM limit of the NMSSM is obtained in the
limit of large $s$ holding $\lam s=\mueff$ fixed and $\kap s$
fixed. Thus, we would expect that the red square points would tend to
have
small $\lam$ and $\kap$.  A later plot will shows this tendency.

\begin{figure}[ht!]

  \centerline{\includegraphics[width=2.4in,angle=90]{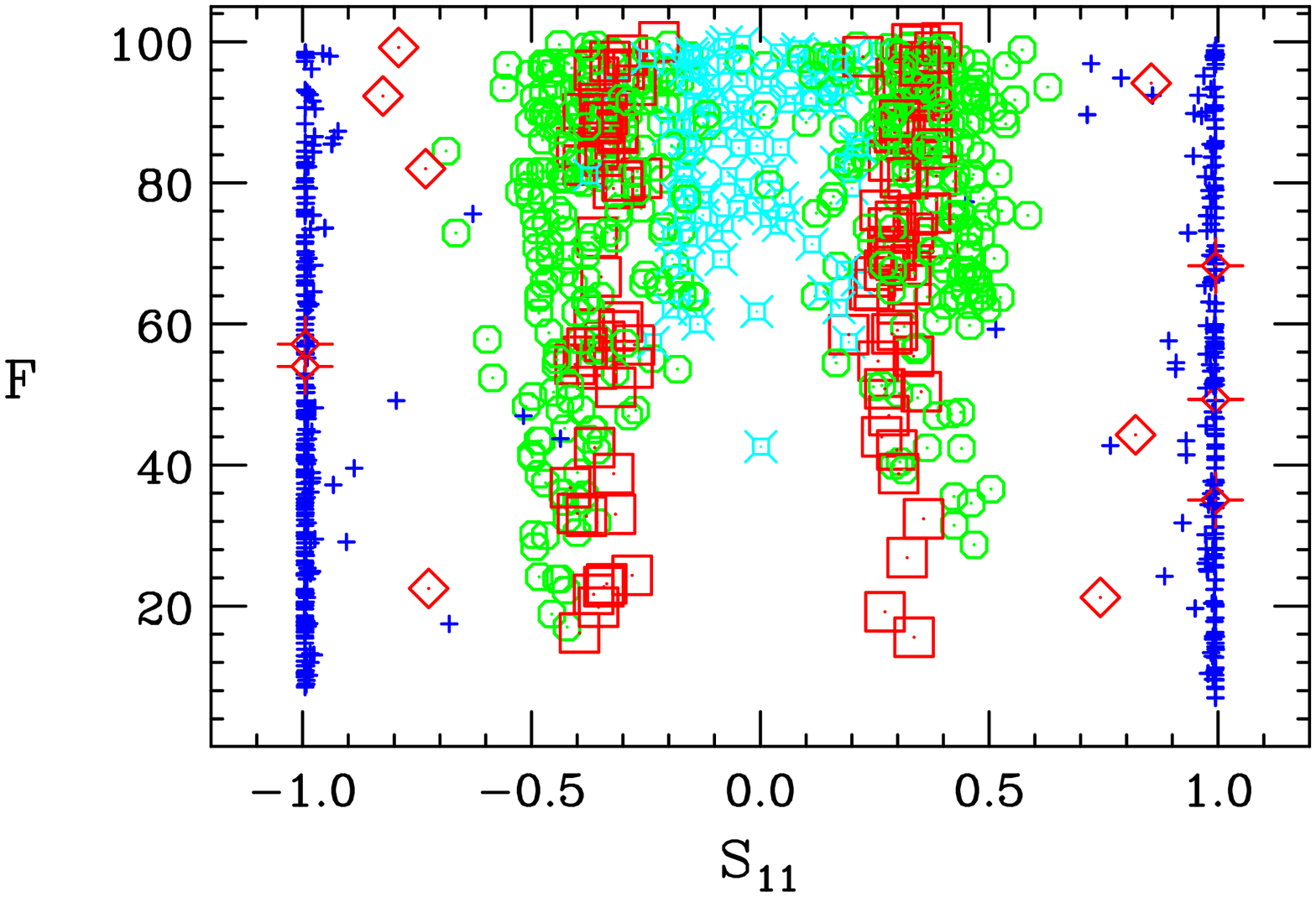}}
  \vspace*{.1in}
  \centerline{\includegraphics[width=2.4in,angle=90]{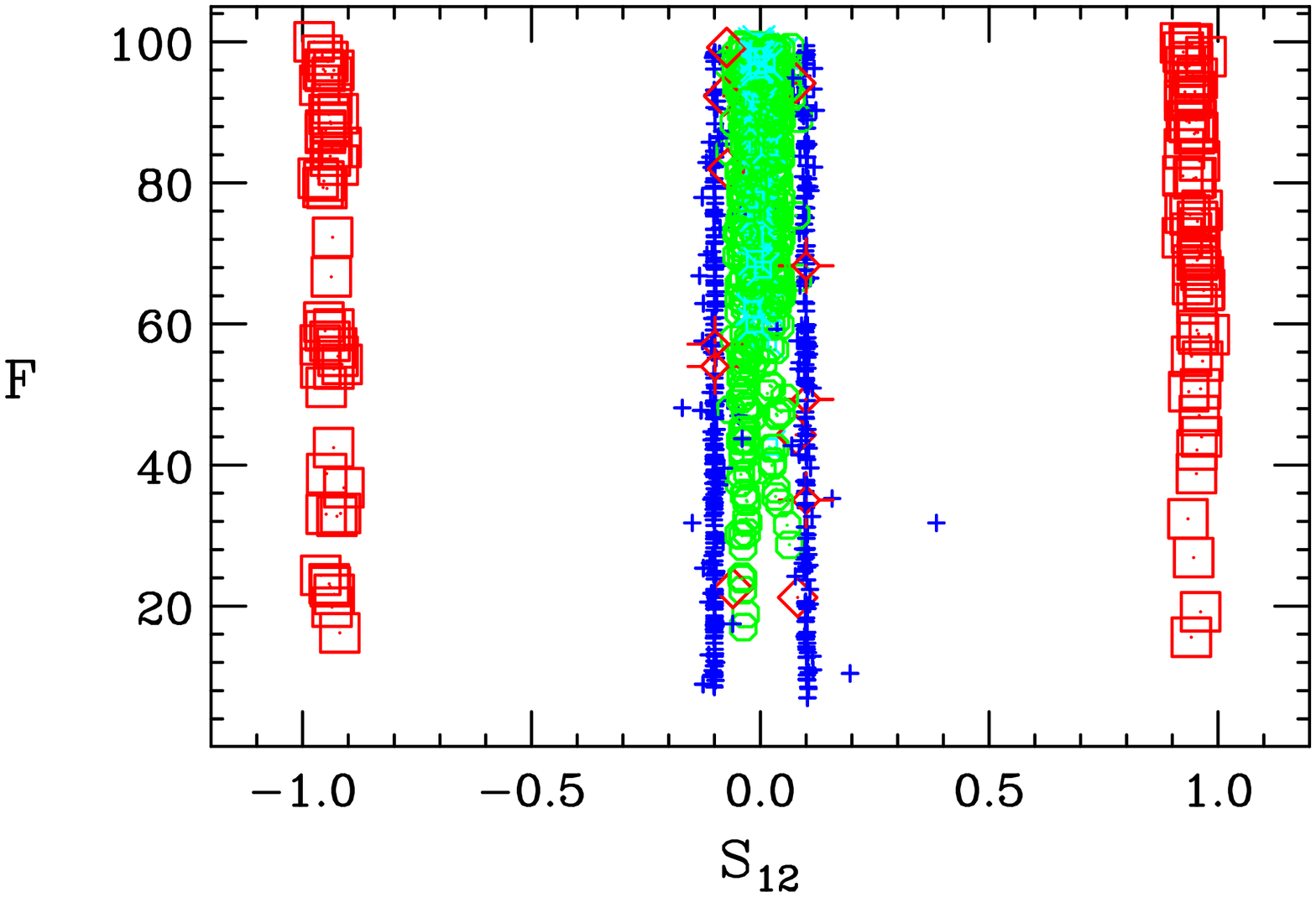}}
  \vspace*{.1in}
  \centerline{\includegraphics[width=2.4in,angle=90]{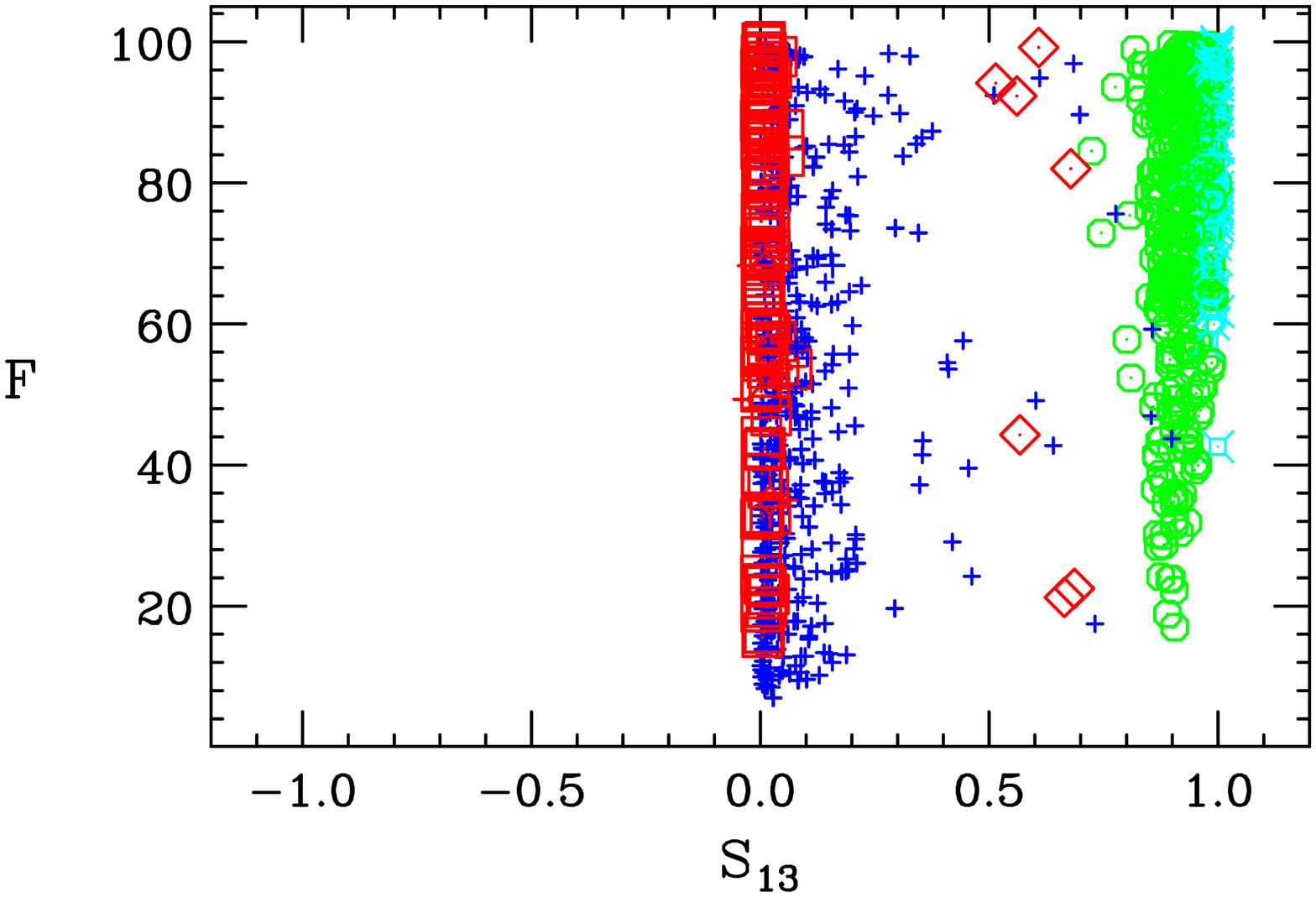}}
\caption{ For fixed $M_{1,2,3}(\mz)=100,200,300\gev$
and $\tanb=10$ we plot: $F$ vs. $S_{11}$ (top); $F$ vs. $S_{12}$
(middle);  and $F$
vs. $S_{13}$ (bottom).  Notation and conventions as in Fig.~\ref{mixplots1}.}
\label{fvssij}
\vspace*{-.1in}
\end{figure}

\begin{figure}[ht!]

  \centerline{\includegraphics[width=2.4in,angle=90]{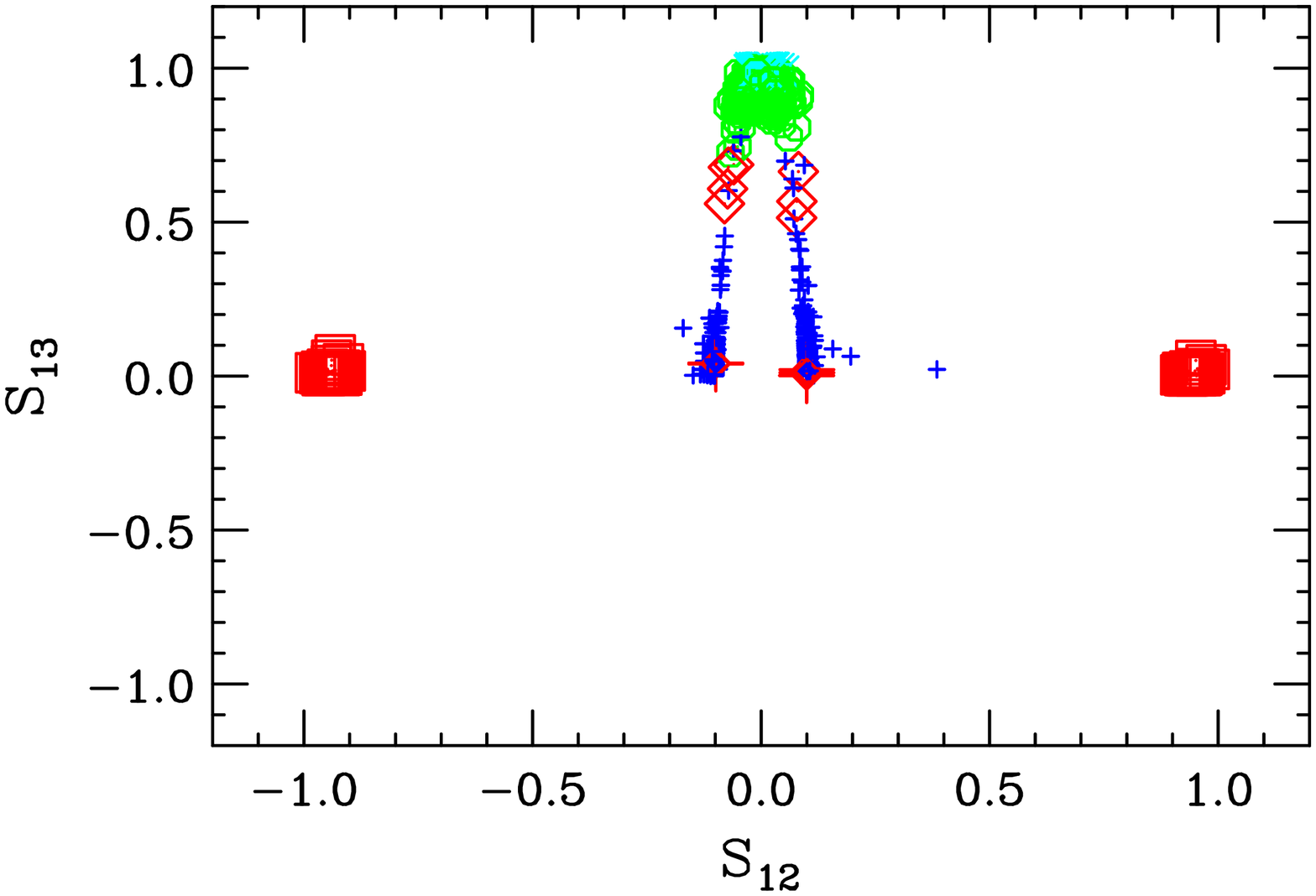}}
  \vspace*{.1in}
  \centerline{\includegraphics[width=2.4in,angle=90]{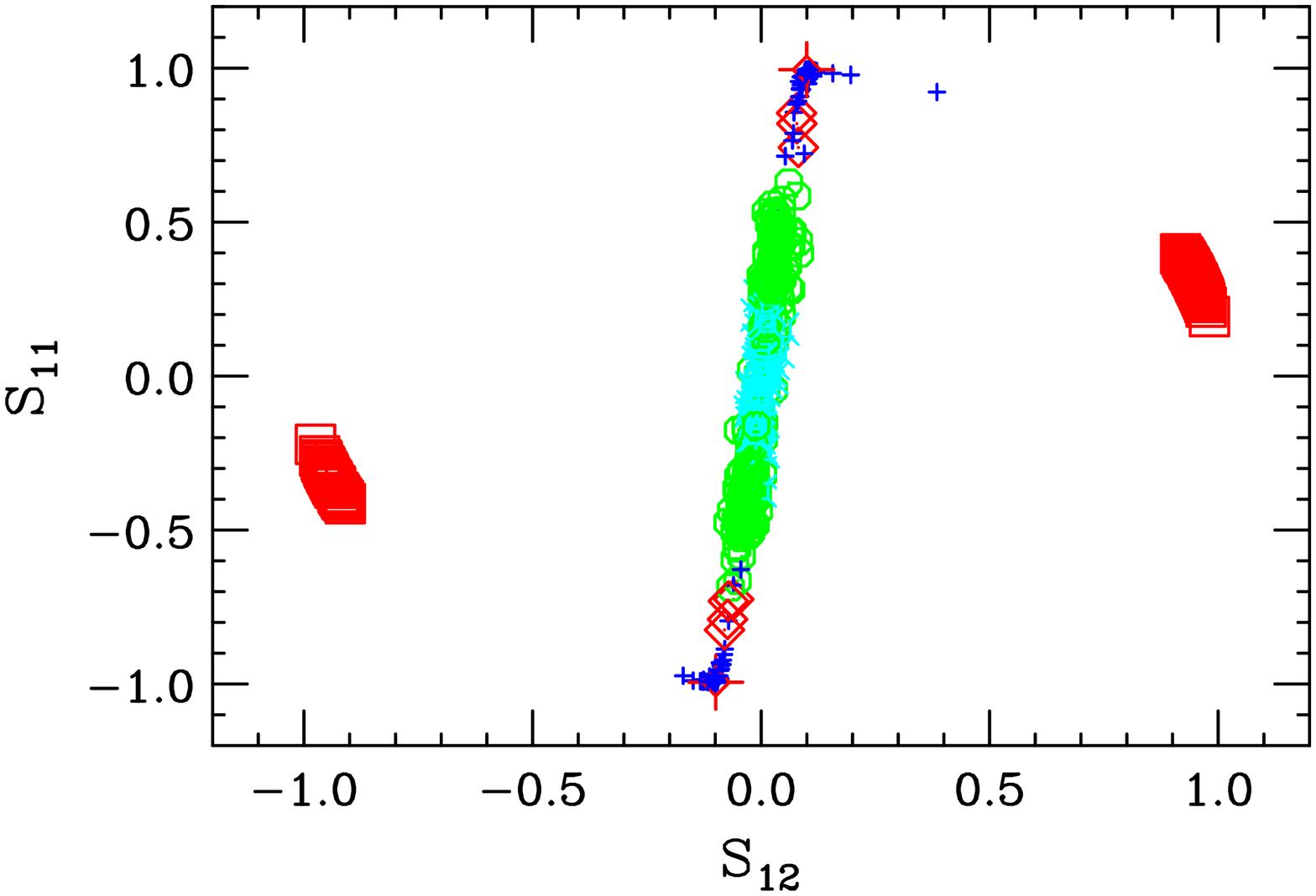}}
  \vspace*{.1in}
  \centerline{\includegraphics[width=2.4in,angle=90]{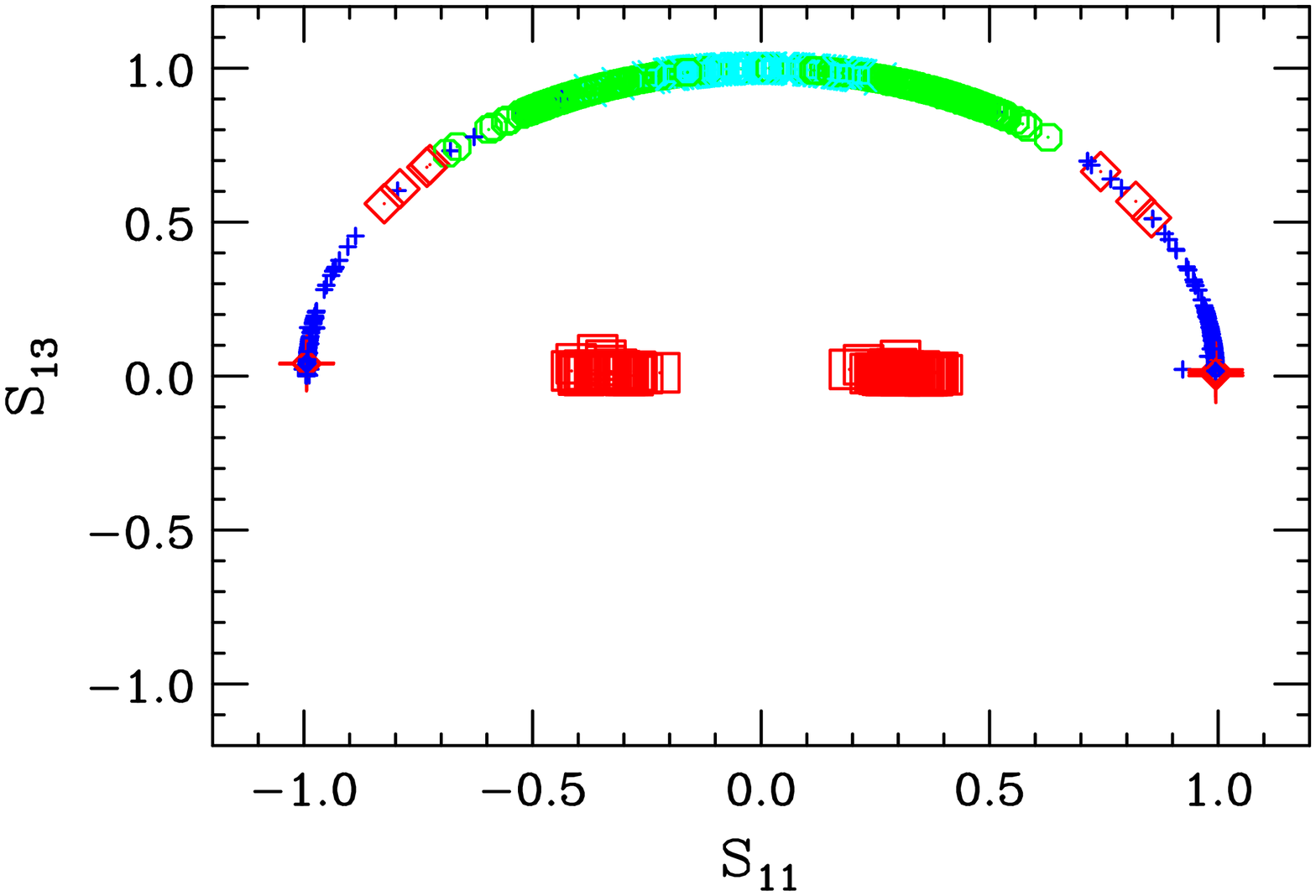}}
\caption{ For fixed $M_{1,2,3}(\mz)=100,200,300\gev$
and $\tanb=10$ we plot: $S_{13}$ vs. $S_{12}$ (top) and $S_{11}$
vs. $S_{12}$ (middle) and $S_{13}$ vs. $S_{11}$ (bottom).  Notation and conventions as in Fig.~\ref{mixplots1}.
}
\label{corrs1}
\vspace*{-.1in}
\end{figure}

\begin{figure}[ht!]
  \centerline{\includegraphics[width=2.4in,angle=90]{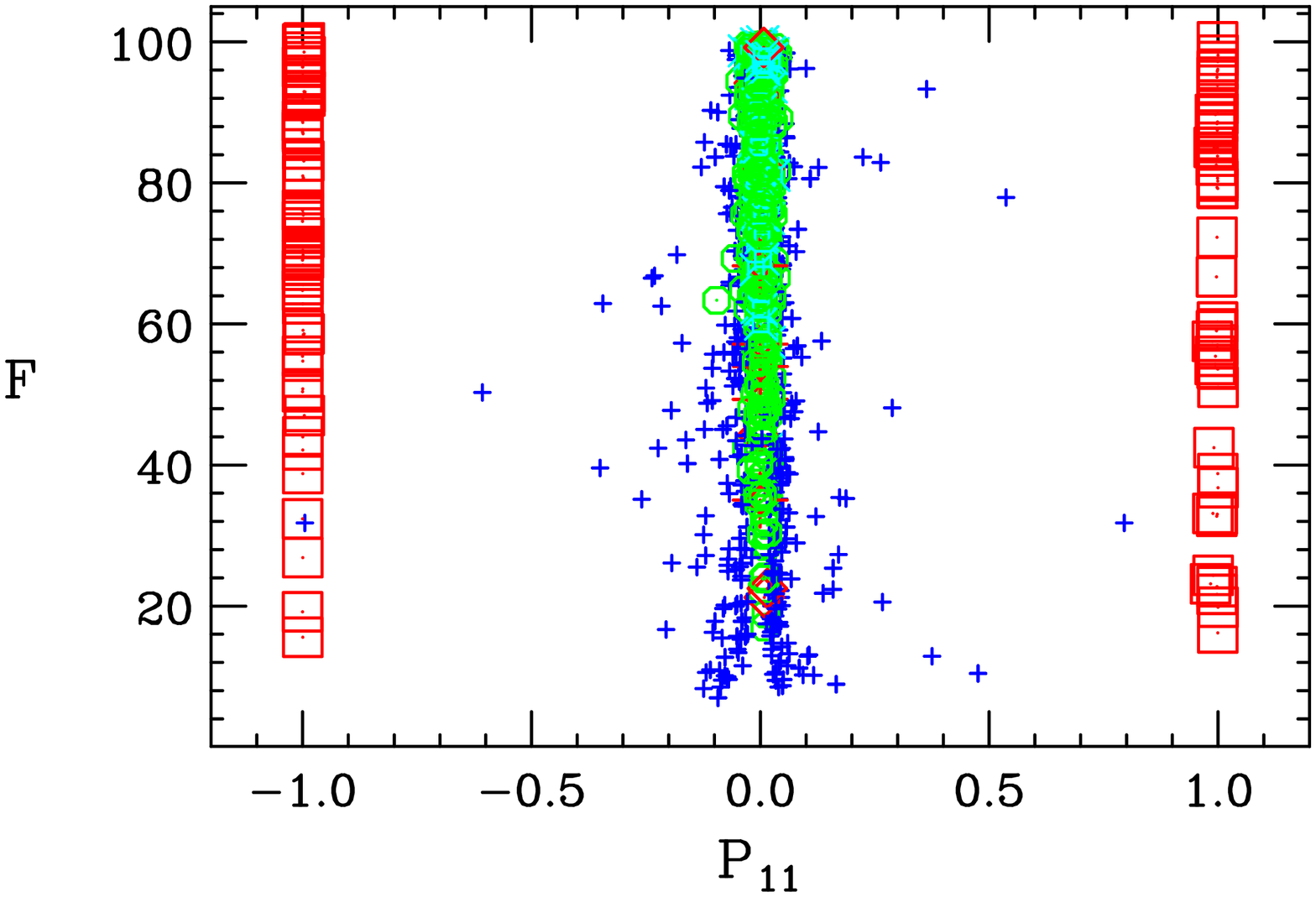}}
  \centerline{\includegraphics[width=2.4in,angle=90]{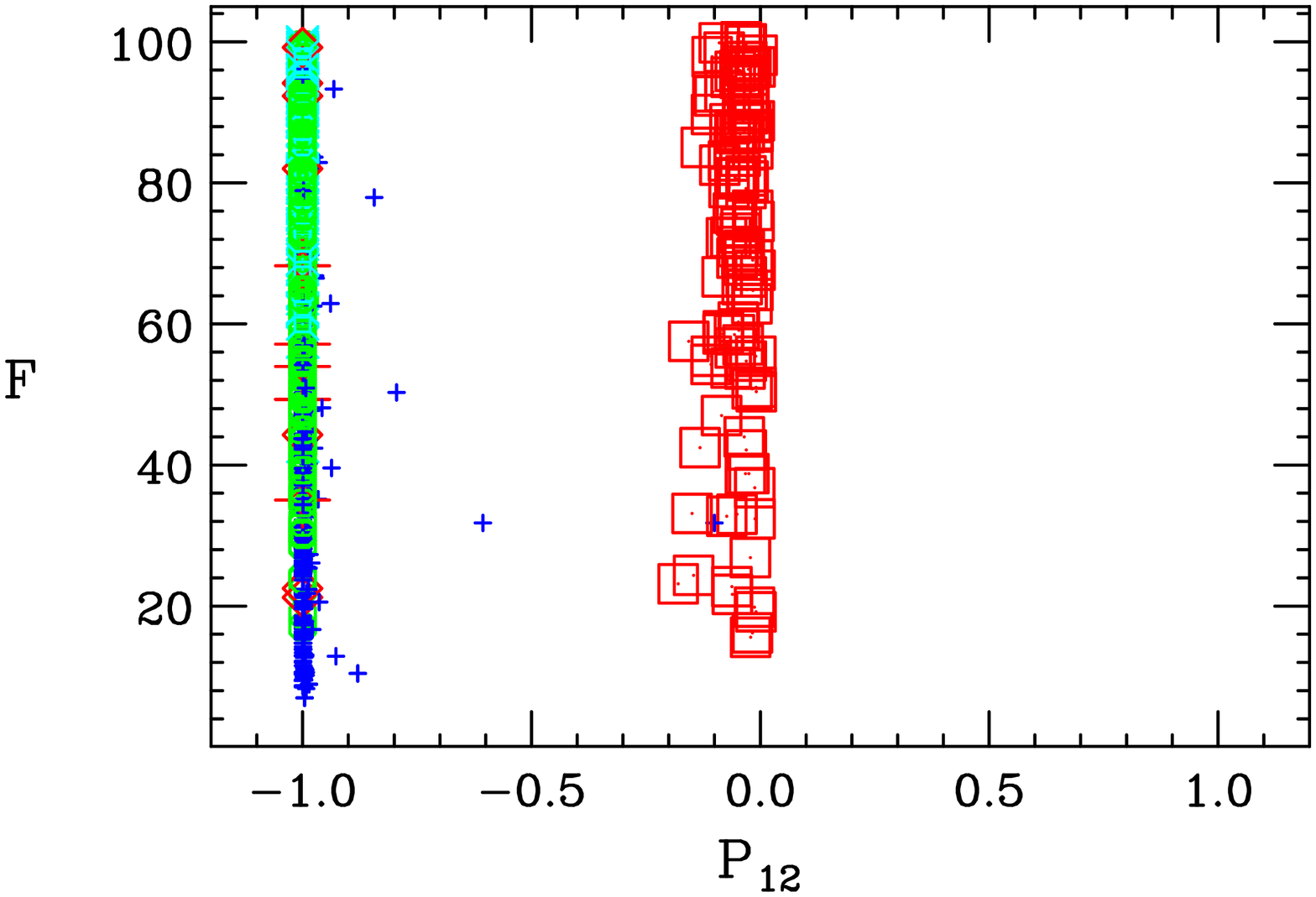}}
  \centerline{\includegraphics[width=2.4in,angle=90]{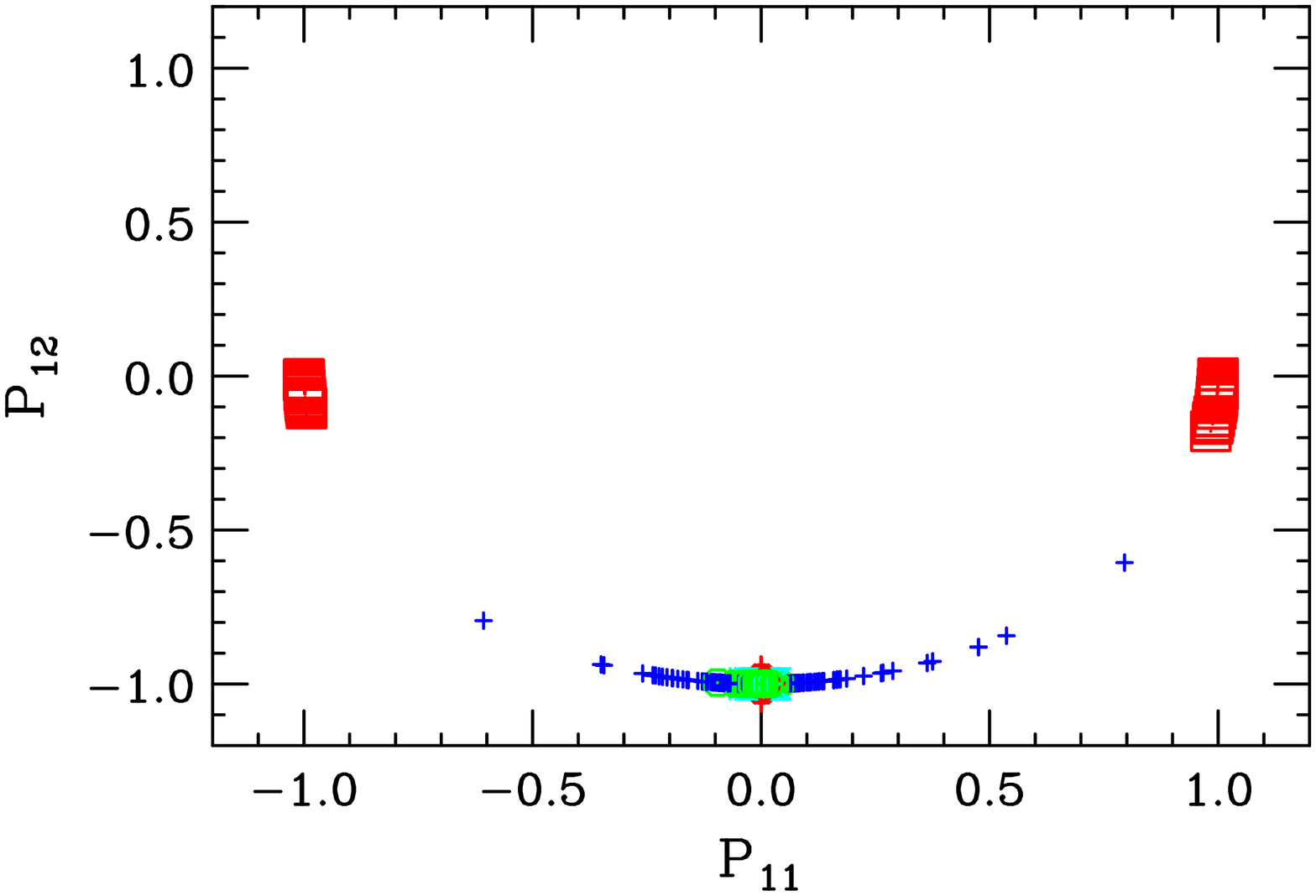}}
\caption{For fixed $M_{1,2,3}(\mz)=100,200,300\gev$
and $\tanb=10$ we plot: $F$ vs. $P_{11}$ (top) and $F$
vs. $P_{12}$ (middle). The bottom plot shows $P_{12}$
vs. $P_{11}$.  Notation and conventions as in Fig.~\ref{mixplots1}. }
\label{corrs2}
\vspace*{-.1in}
\end{figure}

\begin{figure}[ht!]
  \centerline{\includegraphics[width=2.4in,angle=90]{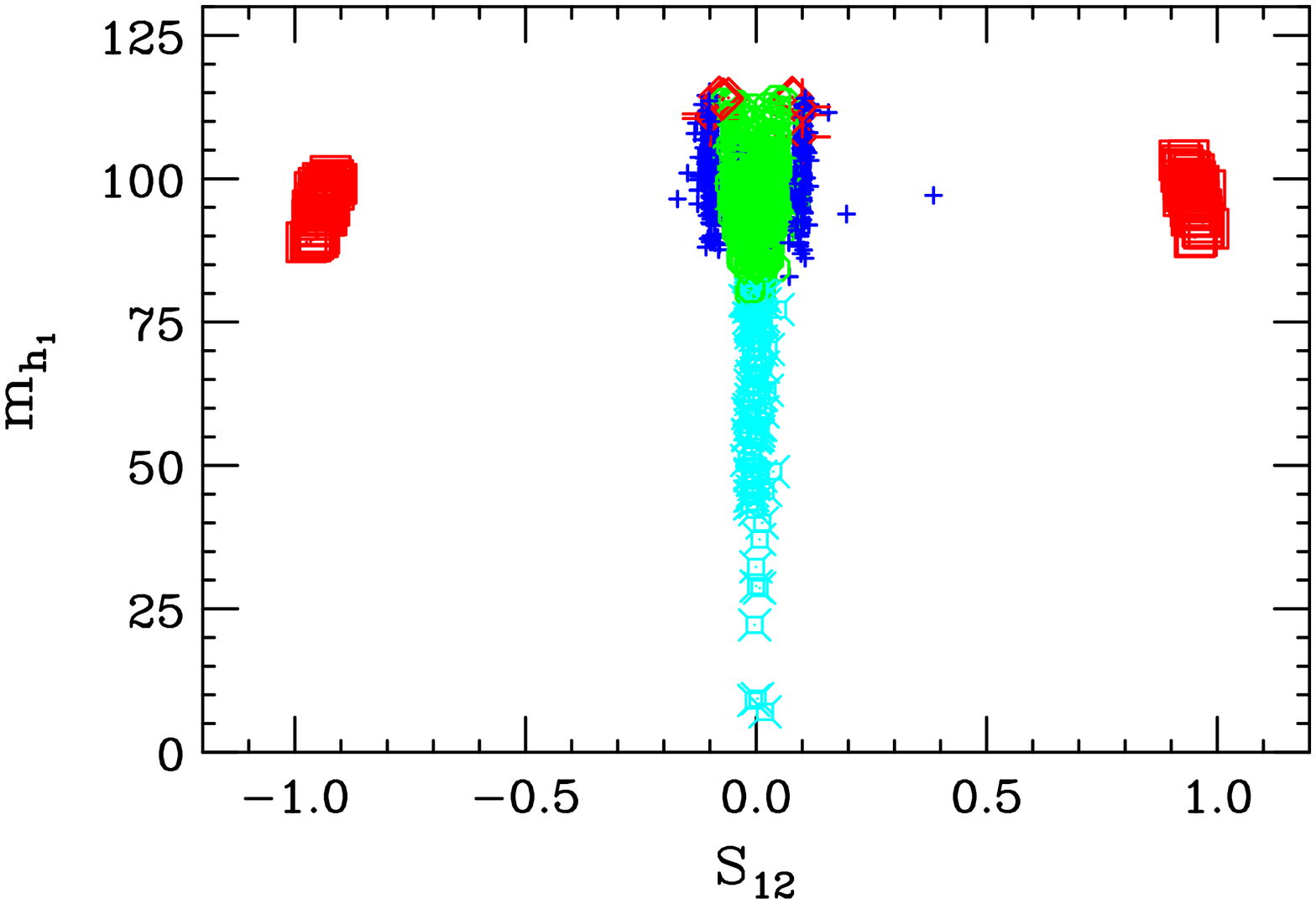}}
\caption{For fixed $M_{1,2,3}(\mz)=100,200,300\gev$
and $\tanb=10$ we plot: $\mhi$ vs. $S_{12}$. Notation and conventions as in Fig.~\ref{mixplots1}.
 }
\label{mh1vss12}
\vspace*{-.1in}
\end{figure}

The various features of all categories of points are illustrated in
detail in a series of figures.  Correlations between $\mhi$ and
$\mhii$ and between $\mhi$ and $\mai$ for the points shown in
Fig.~\ref{mixplots1} are shown in Fig.~\ref{mhimhii}.  Details regarding
$C_V^2(\hi)$ and $C_V^2(\hii)$ are shown in
Fig.~\ref{mixplots2}. The
above-described correlations involving the compositions of the $\hi$
and $\ai$ are made apparent in the plots of
Figs.~\ref{fvssij},~\ref{corrs1}~and~\ref{corrs2}. Fig.~\ref{mh1vss12}
shows $\mhi$ as a function of the $S_{12}$ composition.

\begin{figure}[h!]
  \centerline{\includegraphics[width=2.4in,angle=90]{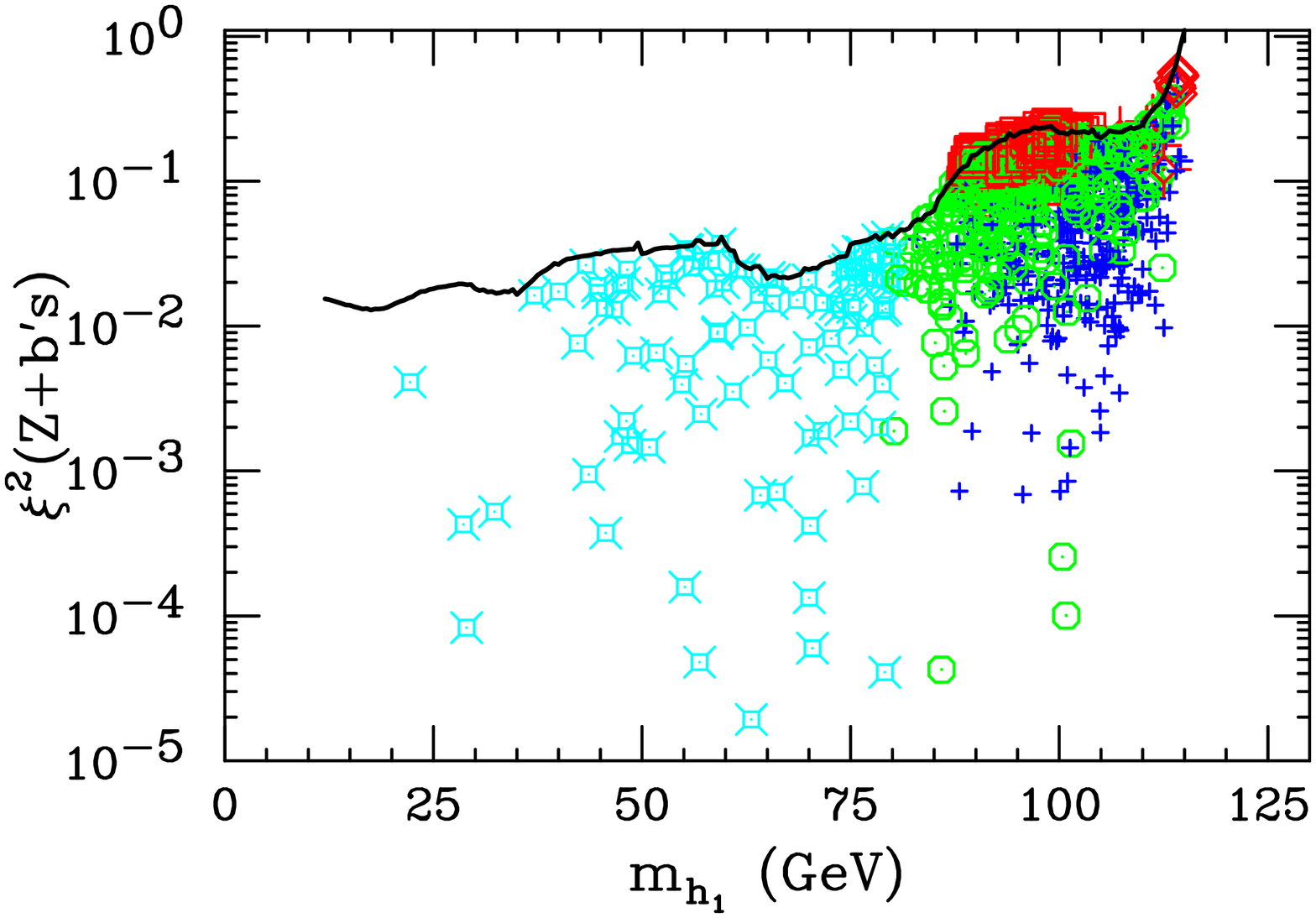}}
\caption{ For fixed $M_{1,2,3}(\mz)=100,200,300\gev$
and $\tanb=10$ we plot $\xi^2(Z+b's)(\hi)$ vs. $\mhi$
for the points from our scan. 
Also shown by the solid line is the approximate LEP experimental
limit. Note the dip in this limit from about $60\gev$ to about
$80\gev$
that cuts away scan points that would have survived with a slightly
less severe limit.
Notation and conventions as in Fig.~\ref{mixplots1}. }
\label{xisqh1}
\vspace*{-.1in}
\end{figure}

Fig.~\ref{xisqh1} shows the LEP limit on $\xi^2(Z+b's)$ in comparison
to the values for the points in our scan.  The plain (red) diamond and
square points hug the LEP limit. The precipitous decline in the
$\xi^2(Z+b's)$ limit as one passes below $\mhi\sim 80\gev$ means that
only points with $F\gsim 40-50$ are found in this region. As noted
earlier, the limit imposed on $\xi^2$ is a bit too severe in cases
where the $\hi$ branching ratio to $b\anti b$ is much smaller than
that to $b\anti b b\anti b$ due to large $\br(\hi\to \ai\ai)$ and
$\mai>2\mb$. However, this does not arise for either the diamond or
square type points, all of which have $\mhi<2\mai$.

%*
%*      PMASS(1-2): CP-odd masses (ordered)
%*
%*      PCOMP(1-2,1-2): Mixing angles: if AB(I) are the bare states,
%*        AB(I) = Im(H1), Im(H2), Im(S), and AM(I) are the mass eigenstates, 
%*        the convention is 
%*        AM(I) = PCOMP(I,1)*(COSBETA*AB(1)+SINBETA*AB(2))
%*                      + PCOMP(I,2)*AB(3)

\begin{figure}[ht!]
  \centerline{\includegraphics[width=2.4in,angle=90]{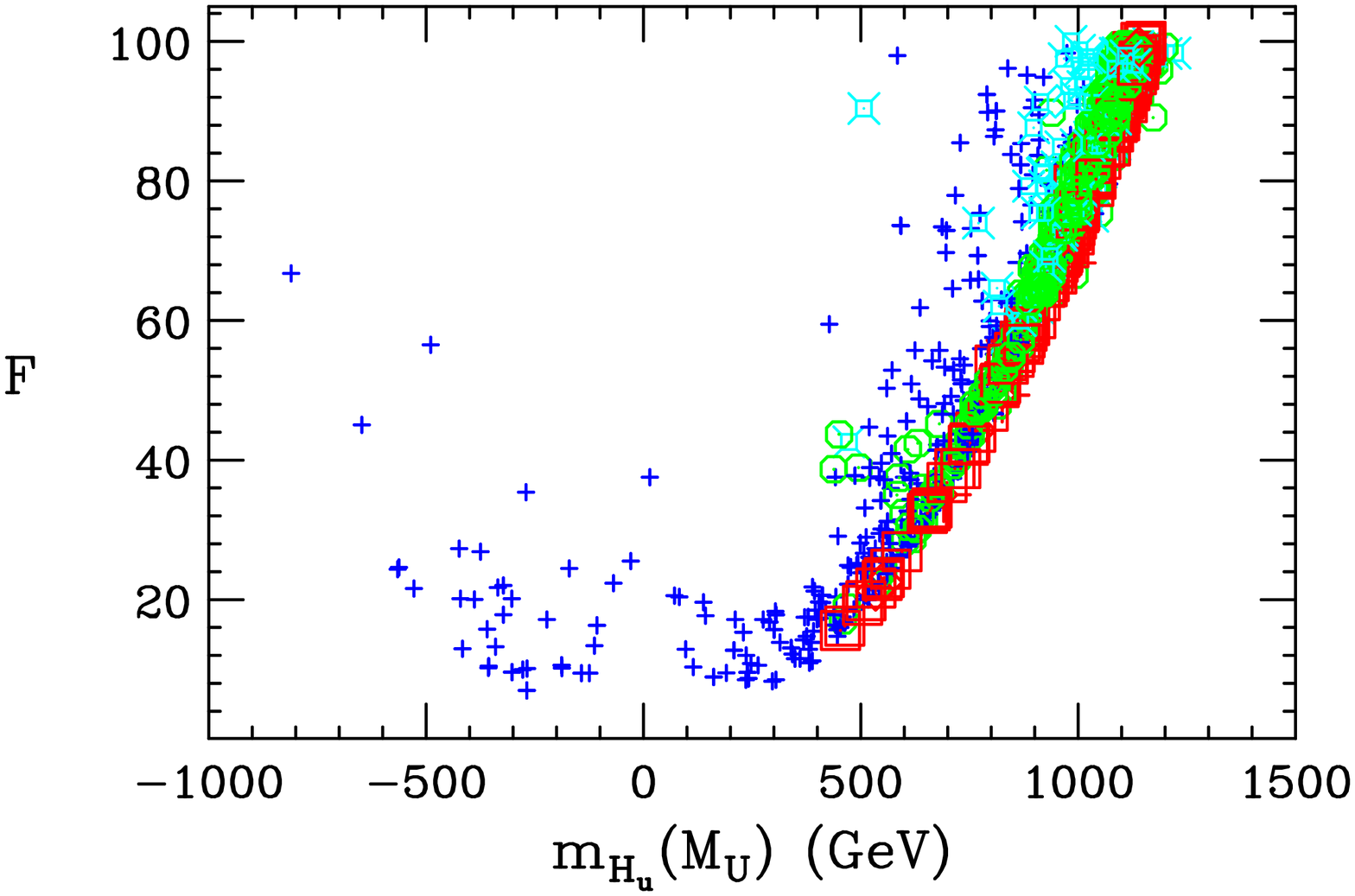}}
  \centerline{\includegraphics[width=2.4in,angle=90]{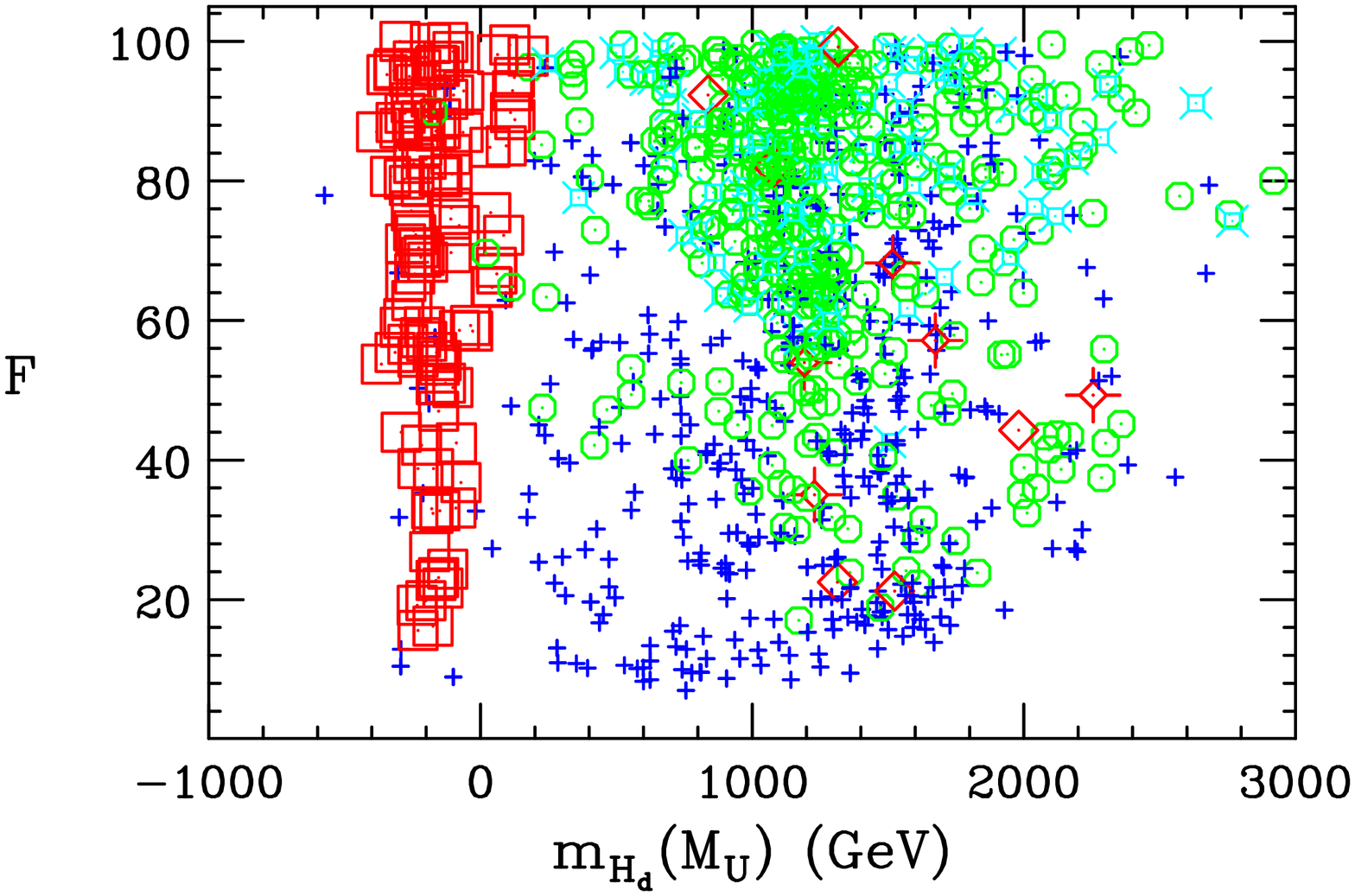}}
  \centerline{\includegraphics[width=2.4in,angle=90]{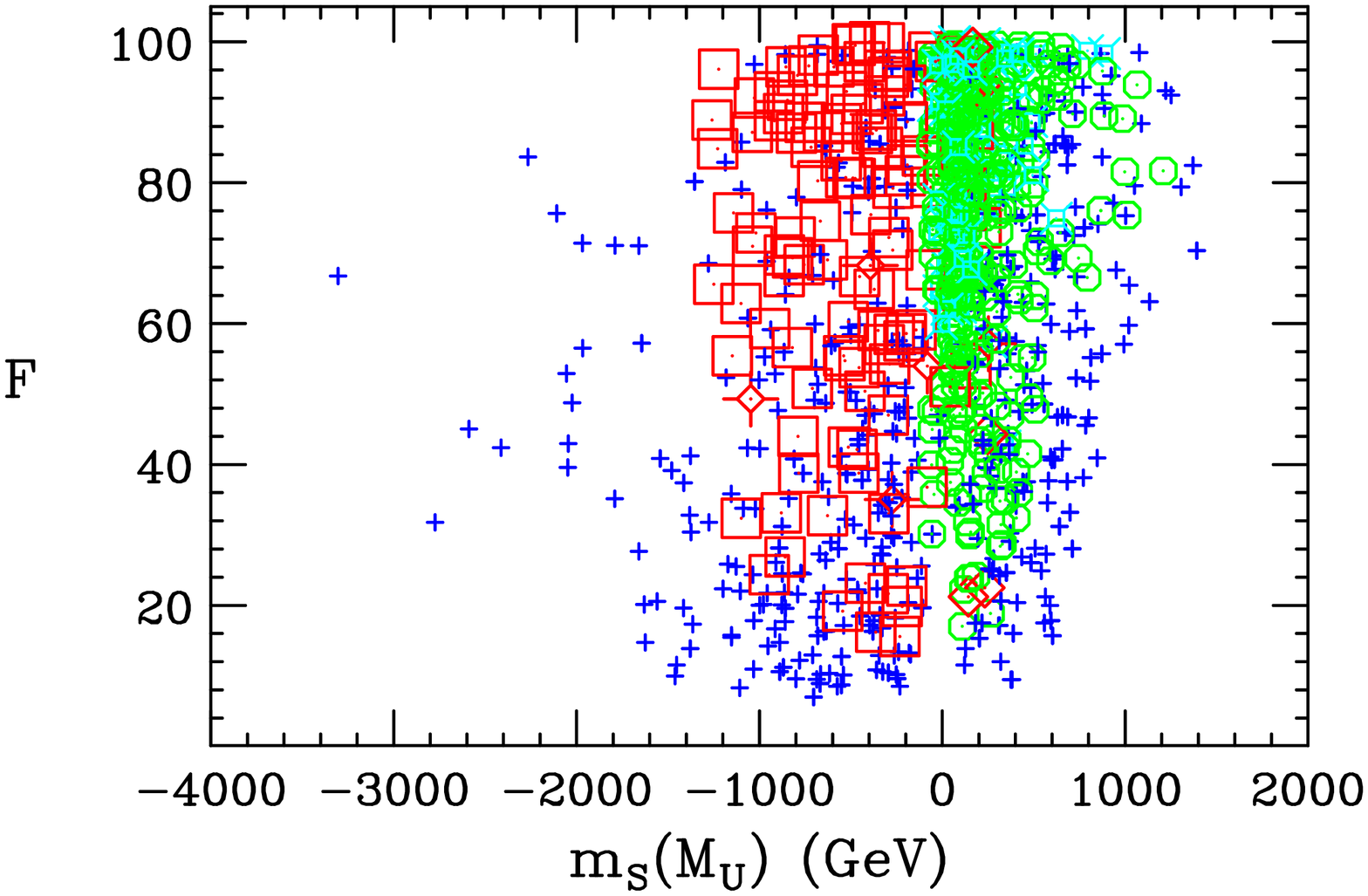}}
\caption{ For fixed $M_{1,2,3}(\mz)=100,200,300\gev$
and $\tanb=10$ we plot $F$ as a function of $m_{H_u}(\mgut)$,
$m_{H_d}(\mgut)$ and $m_S(\mgut)$, where negative $m$ values are
obtained as $-\sqrt{-m^2}$.  Notation and conventions as in Fig.~\ref{mixplots1}. }
\label{mixplots3}
\vspace*{-.1in}
\end{figure}

%\begin{figure}[ht!]
%  \centerline{\includegraphics[width=2.4in,angle=90]{lowfallok_write_specialmix_47.ps}}
%  \centerline{\includegraphics[width=2.4in,angle=90]{lowfallok_write_specialmix_29.ps}}
%  \centerline{\includegraphics[width=2.4in,angle=90]{lowfallok_write_specialmix_48.ps}}
%\caption{ For fixed $M_{1,2,3}(\mz)=100,200,300\gev$
%and $\tanb=10$ we plot $m_S(\mgut)$  as a function of
%$m_{H_u}(\mgut)$, $m_S(\mgut)$ as a function of $m_{H_d}(\mgut)$ and 
%$m_{H_d}(\mgut)$ as a function of $m_{H_u}(\mgut)$, where negative $m$ values are
%obtained as $-\sqrt{-m^2}$.  Notation and conventions as in Fig.~\ref{mixplots1}. }
%\label{mixplots3a}
%\vspace*{-.1in}
%\end{figure}

\begin{figure}[ht!]
  \centerline{\includegraphics[width=2.4in,angle=90]{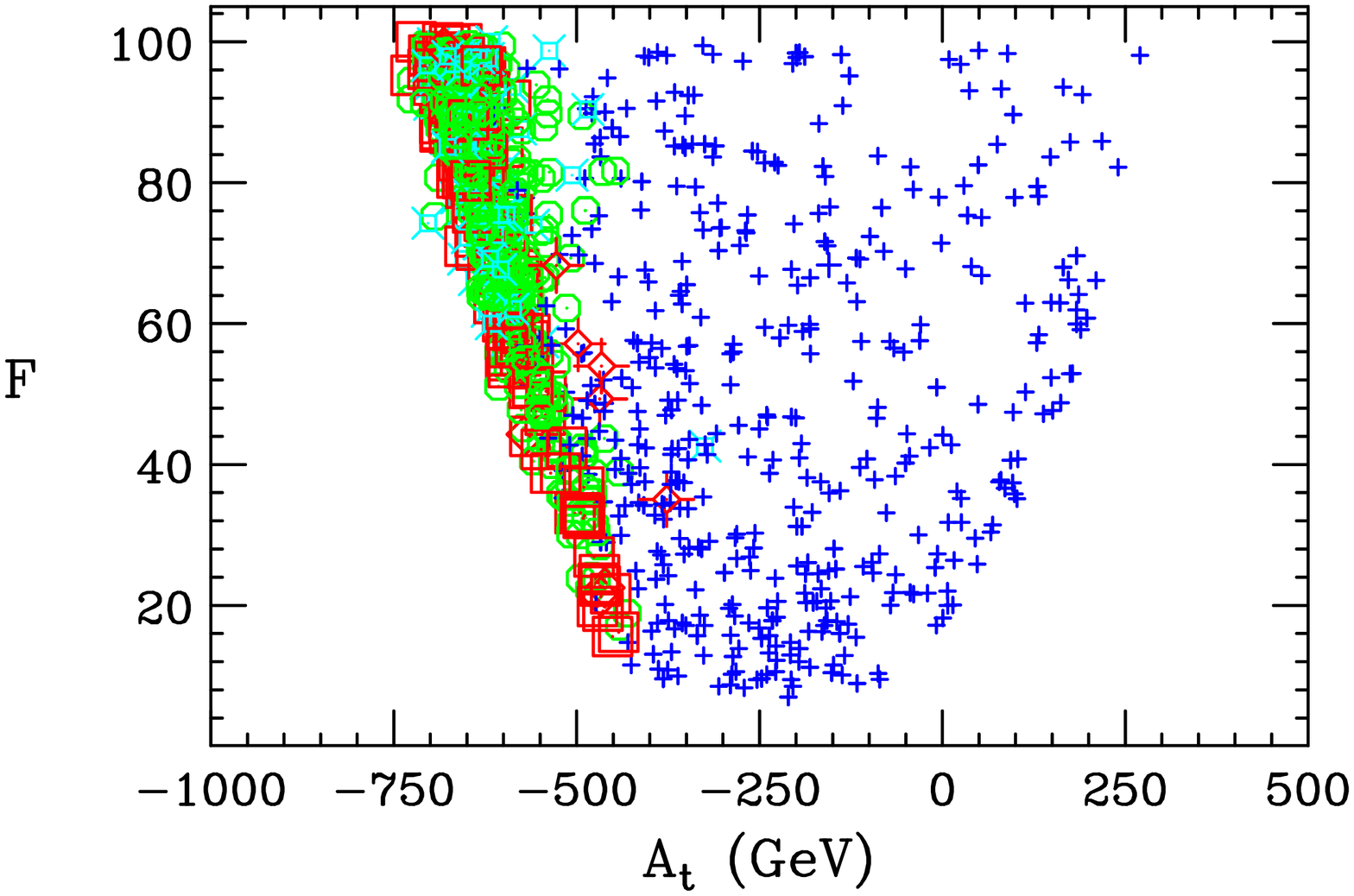}}
  \centerline{\includegraphics[width=2.4in,angle=90]{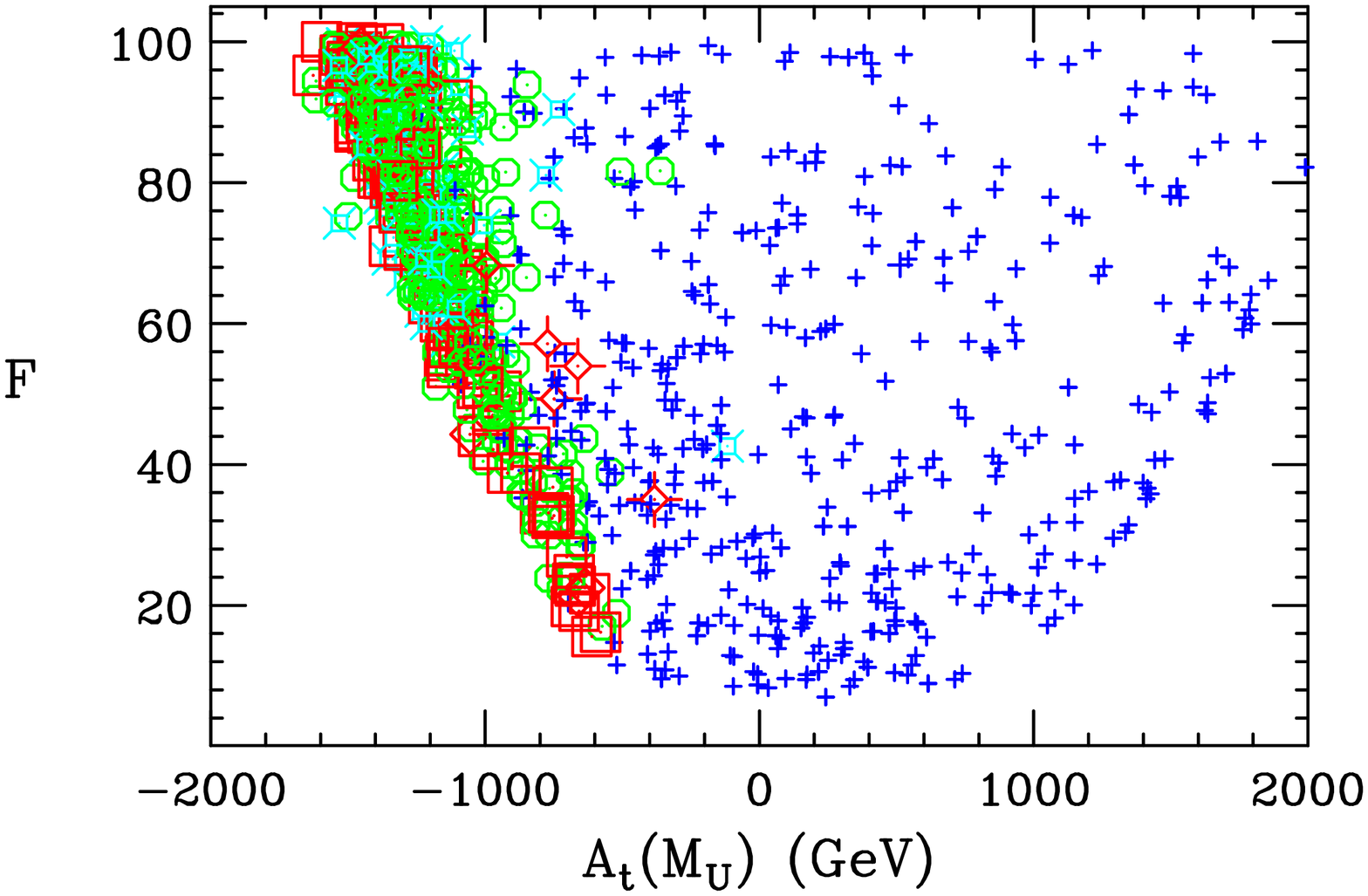}}
\caption{ For fixed $M_{1,2,3}(\mz)=100,200,300\gev$
and $\tanb=10$ we plot $F$ as a function of $A_t$ (at scale $\mz$) and
of $A_t(\mgut)$.  Notation and conventions as in Fig.~\ref{mixplots1}. }
\label{mixplots4}
\vspace*{-.1in}
\end{figure}

Correlations of these scenarios with various GUT-scale parameters are
illuminating.  In Fig.~\ref{mixplots3}, we show $F$ vs.
$m_{H_u}(\mgut)$, $m_{H_d}(\mgut)$ and $m_S(\mgut)$, where negative
$m$ values correspond to $-\sqrt{-m^2}$ with $m^2<0$.  Low fine tuning
is often associated with one or more of the GUT-scale soft Higgs
masses-squared being small. In particular, small
$|m_{H_u}(\mgut)|$ is needed for small $F$. Indeed, once
$|M_{H_U}(\mgut)$ is above $\sim 500\gev$, it is the parameter
$p=\mhusq(\mgut)$ that gives the maximum $F_p$ and $F$ becomes large.
 Note that the MSSM-like
mixed-Higgs scenarios (large plain red squares) require
$m_{H_u}(\mgut) \sim 500\gev$, $m_{H_d}(\mgut)\sim 0$, and modestly
negative $m_S(\mgut)$. The scenarios with large singlet mixing (green
circles and cyan starred-squares) and fairly low $F$ tend to have substantial
$m_{H_u}(\mgut)$ and $m_{H_d}(\mgut)$, but relatively small
$m_S(\mgut)$. The low-$F$ blue $+$ scenarios with a light SM-like $\hi$ are
more spread out in all these parameters, but are also easily obtained if
all the GUT-scale soft Higgs masses-squared are relatively small. 

It is also useful to examine $F$ vs. $A_t$ and
$A_t(\mgut)$ as shown in Fig.~\ref{mixplots4}.  As noted in our
earlier paper, the lowest $F$ (blue) $+$ points require quite small
$A_t(\mgut)$. (Of course, by 'small', we do not mean zero. Typically,
all these parameters have magnitudes given by a scale of order
$100-200\gev$, \ie\ of order $\mz$ itself.) The lowest $F$ values for
the mixed-Higgs scenarios are achieved (as in the MSSM) for negative
$A_t(\mz)\lsim -500\gev$.

\begin{figure}[ht!]
  \centerline{\includegraphics[width=2.4in,angle=90]{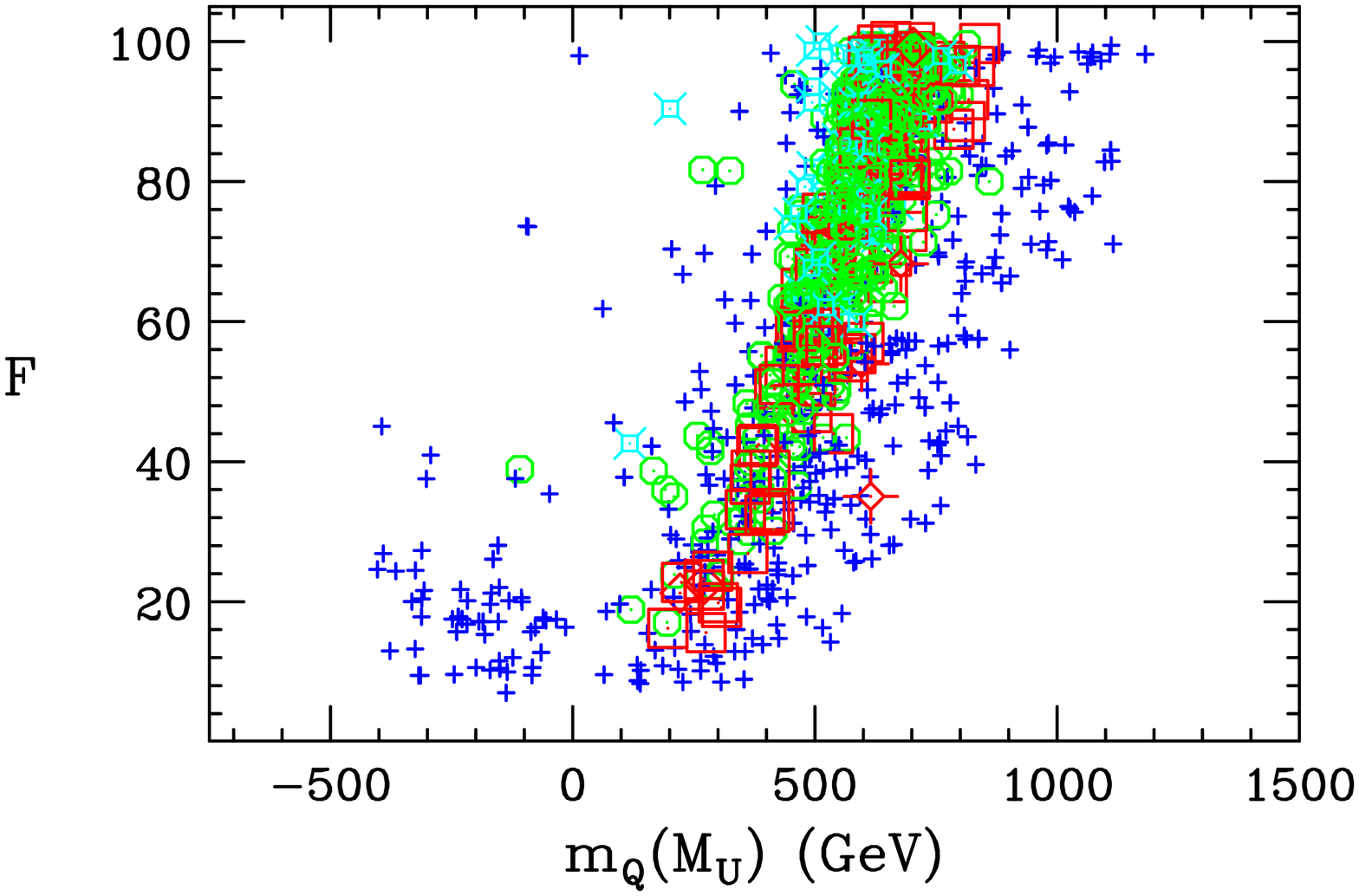}}
  \centerline{\includegraphics[width=2.4in,angle=90]{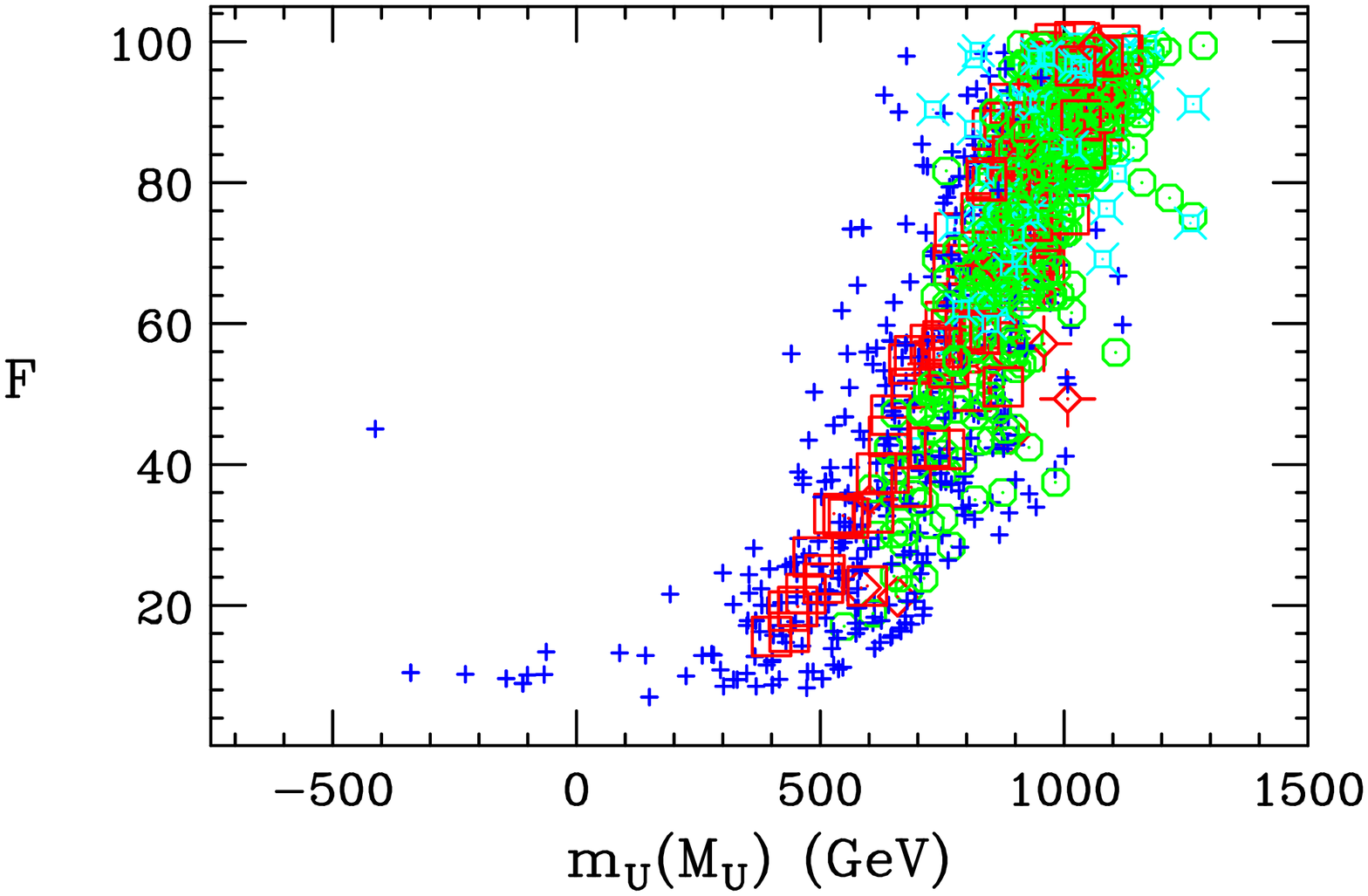}}
  \centerline{\includegraphics[width=2.4in,angle=90]{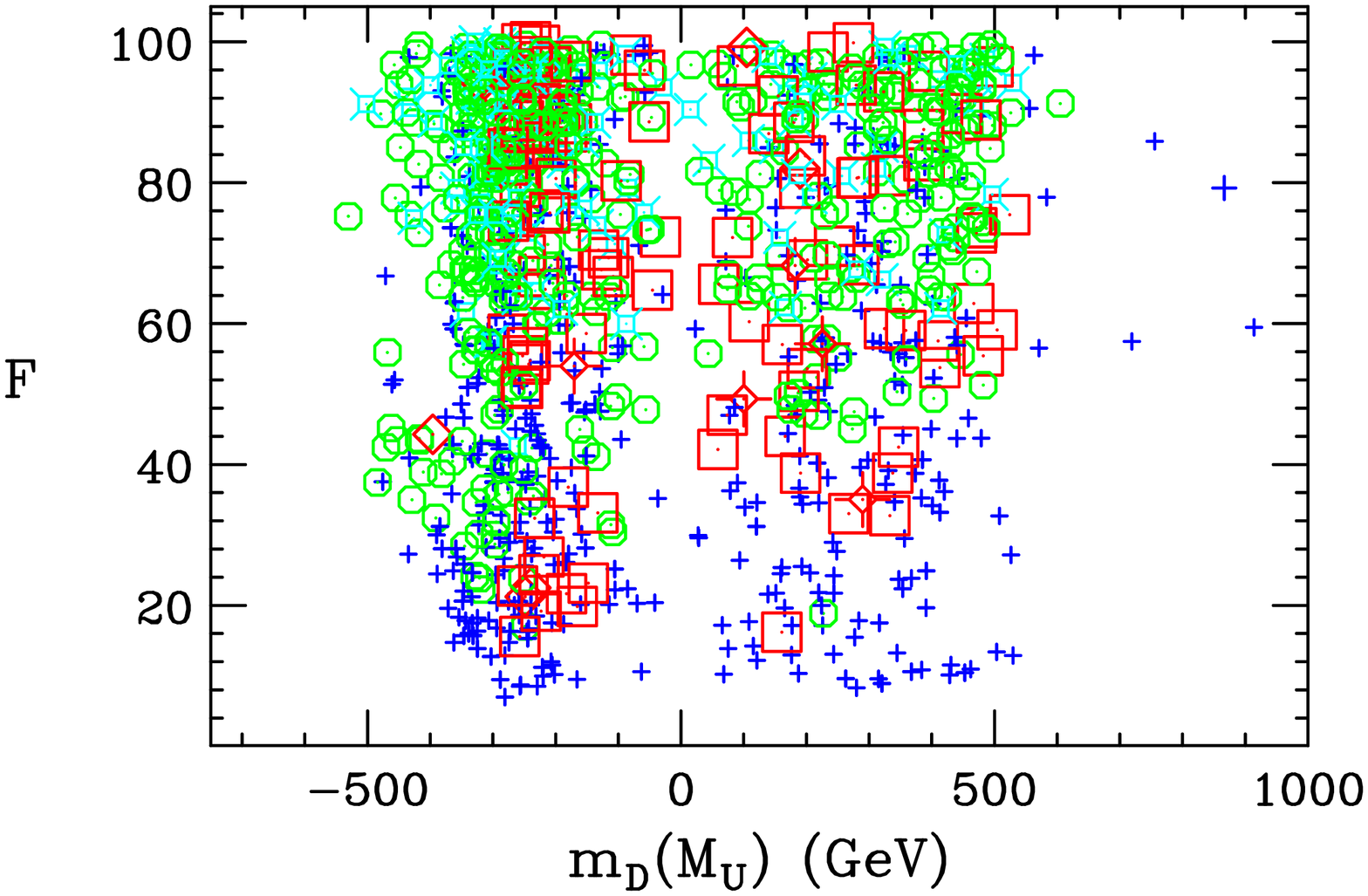}}
\caption{ For fixed $M_{1,2,3}(\mz)=100,200,300\gev$
and $\tanb=10$ we plot $F$ as a function of $m_Q(\mgut)$, $m_U(\mgut)$
and $m_D(\mgut)$. For all three $m$'s, if $m^2>0$ ($m^2<0$)
then $m\equiv \sqrt{m^2}$ ($m\equiv-\sqrt{-m^2}$).
 Notation and conventions as in Fig.~\ref{mixplots1}.}
\label{mixplots5}
\vspace*{-.1in}
\end{figure}

It is also amusing to examine $F$ as a function of $m_Q(\mgut)$,
$m_U(\mgut)$ and $m_D(\mgut)$ for the third generation.  The plots
appear in Fig.~\ref{mixplots5}. We observe that the smallest $F$
(blue) $+$ points require $m_Q(\mgut)$, $m_U(\mgut)$ and $m_D(\mgut)$
values of order a few hundred $\gev$, while larger values are
typically required for the various mixed-Higgs scenarios.

\begin{figure}[t!]
  \centerline{\includegraphics[width=2.4in,angle=90]{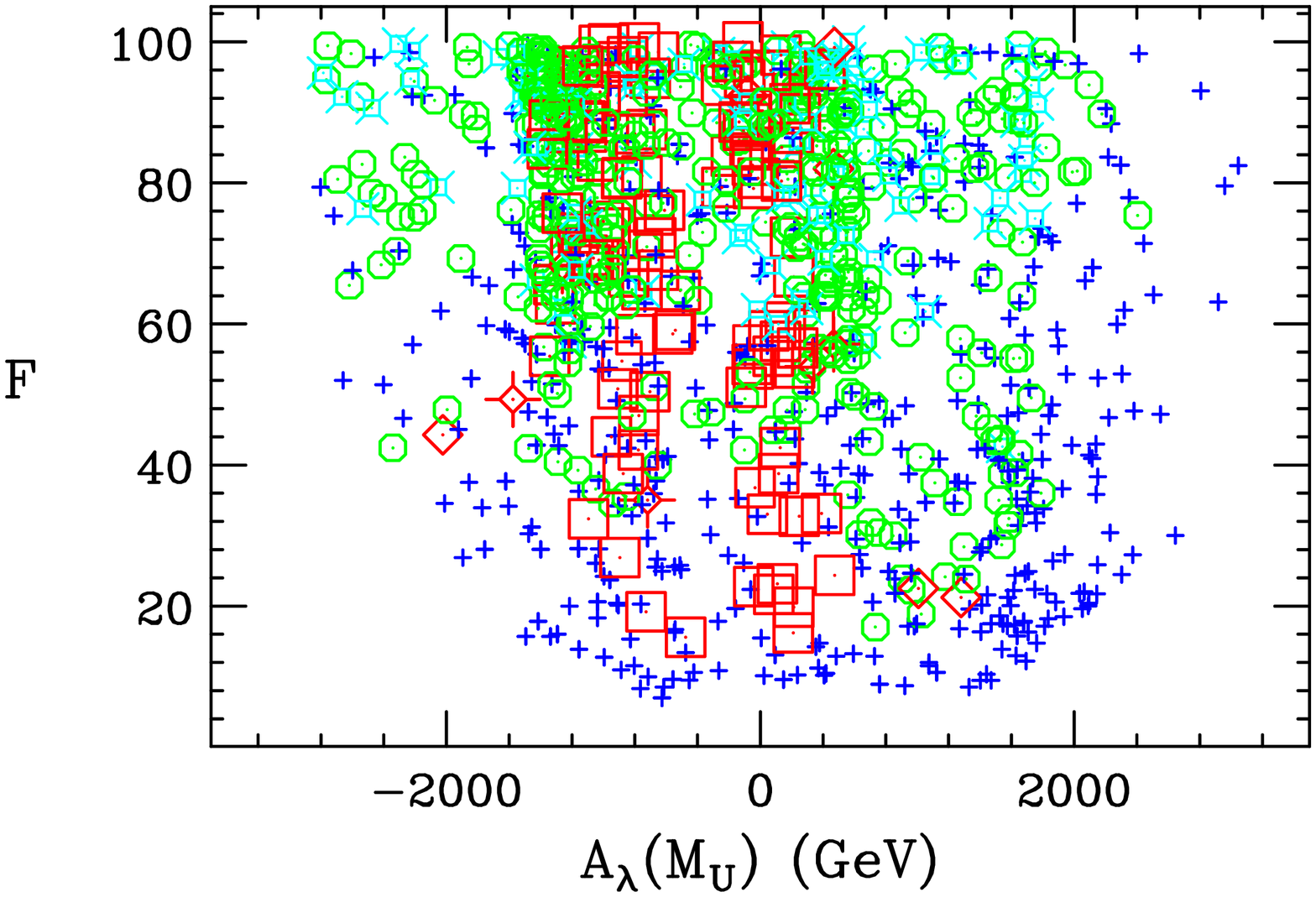}}
  \centerline{\includegraphics[width=2.4in,angle=90]{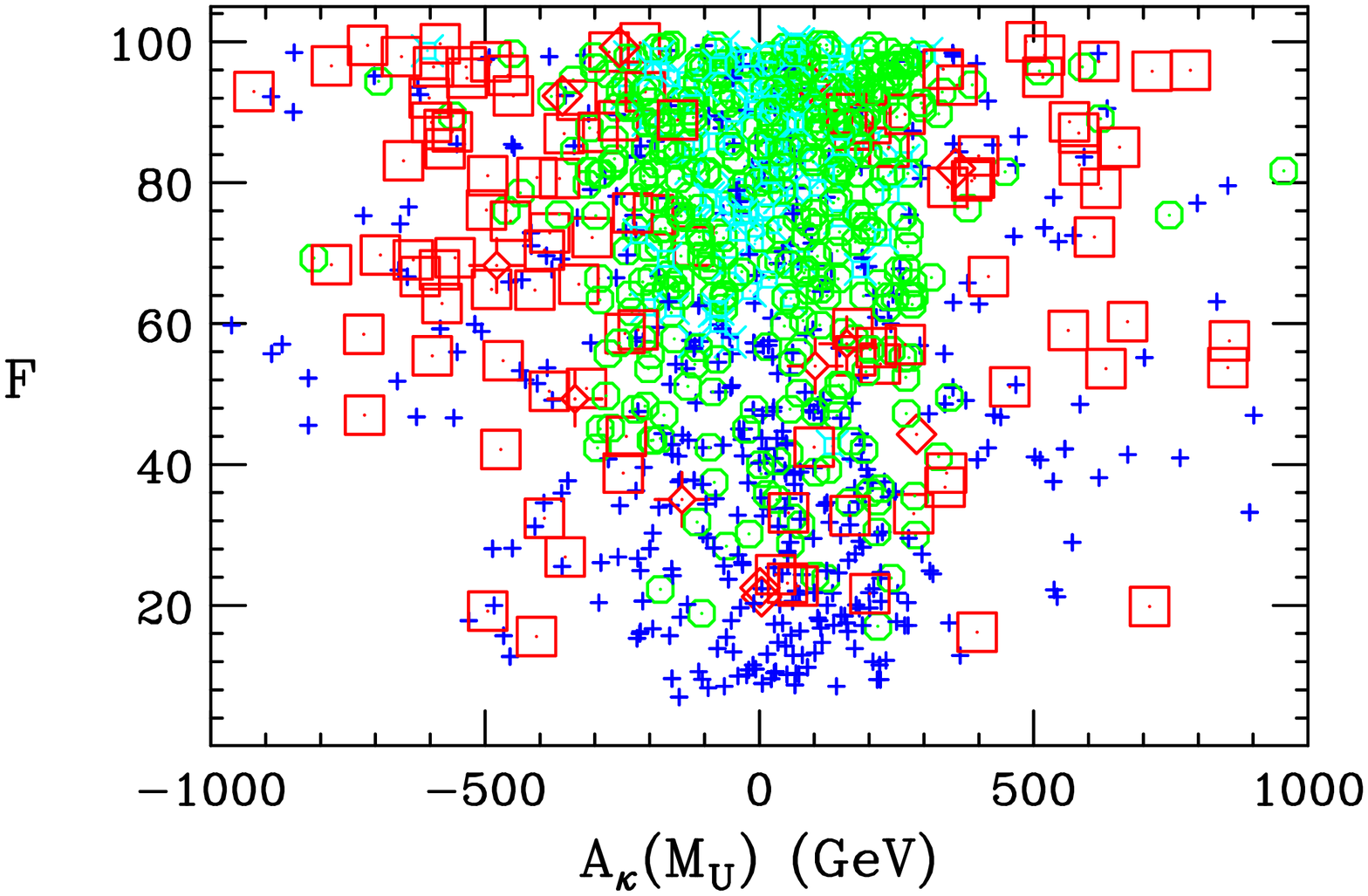}}
\vspace*{-.15in}
\caption{ For fixed $M_{1,2,3}(\mz)=100,200,300\gev$
and $\tanb=10$ we plot $F$ as a function of $\alam(\mgut)$ and
$\akap(\mgut)$. Notation and conventions as in Fig.~\ref{mixplots1}.}
\label{mixplots6}

\end{figure}

\begin{figure}[h!]
  \centerline{\includegraphics[width=2.4in,angle=90]{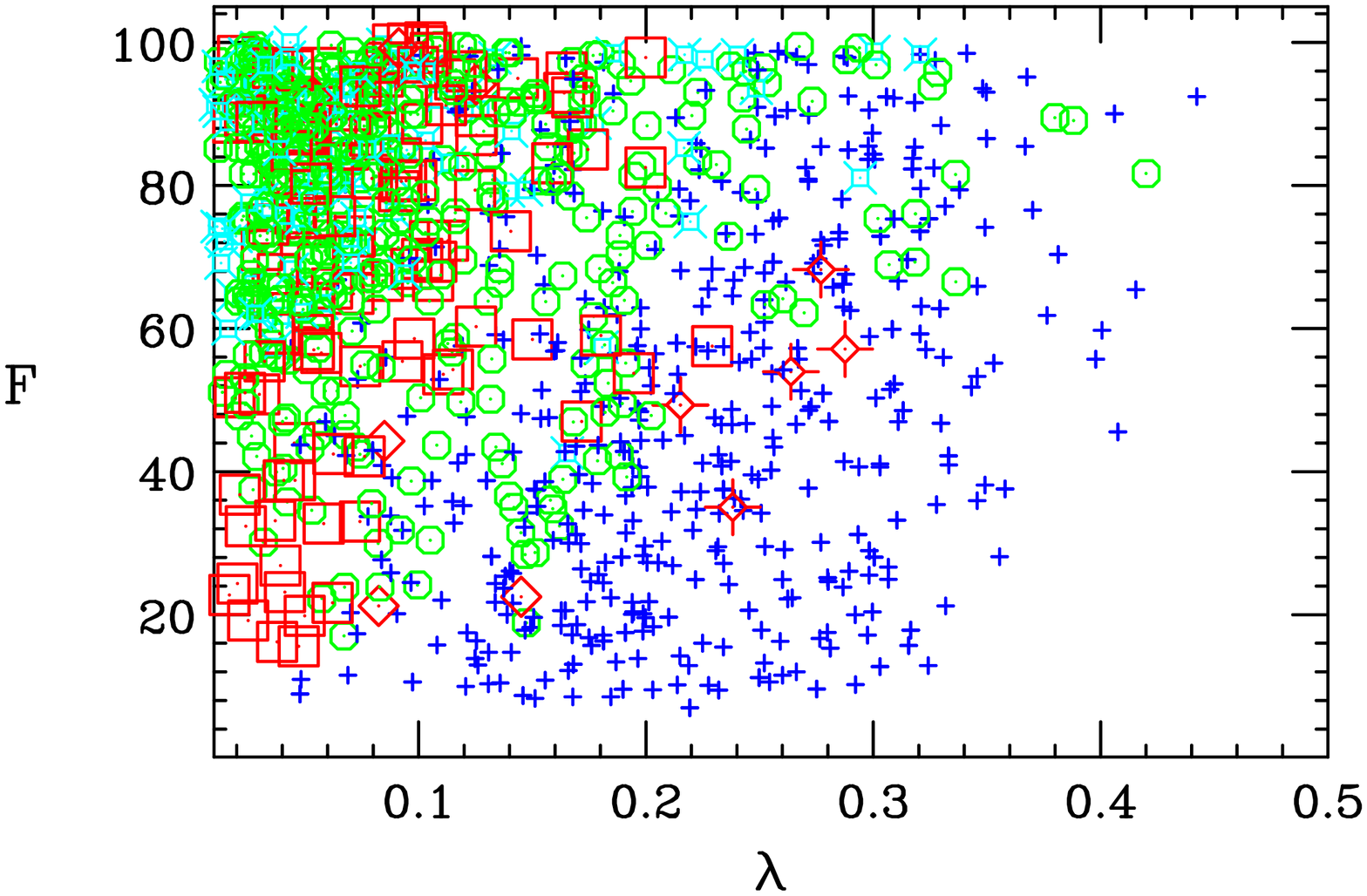}}
  \centerline{\includegraphics[width=2.4in,angle=90]{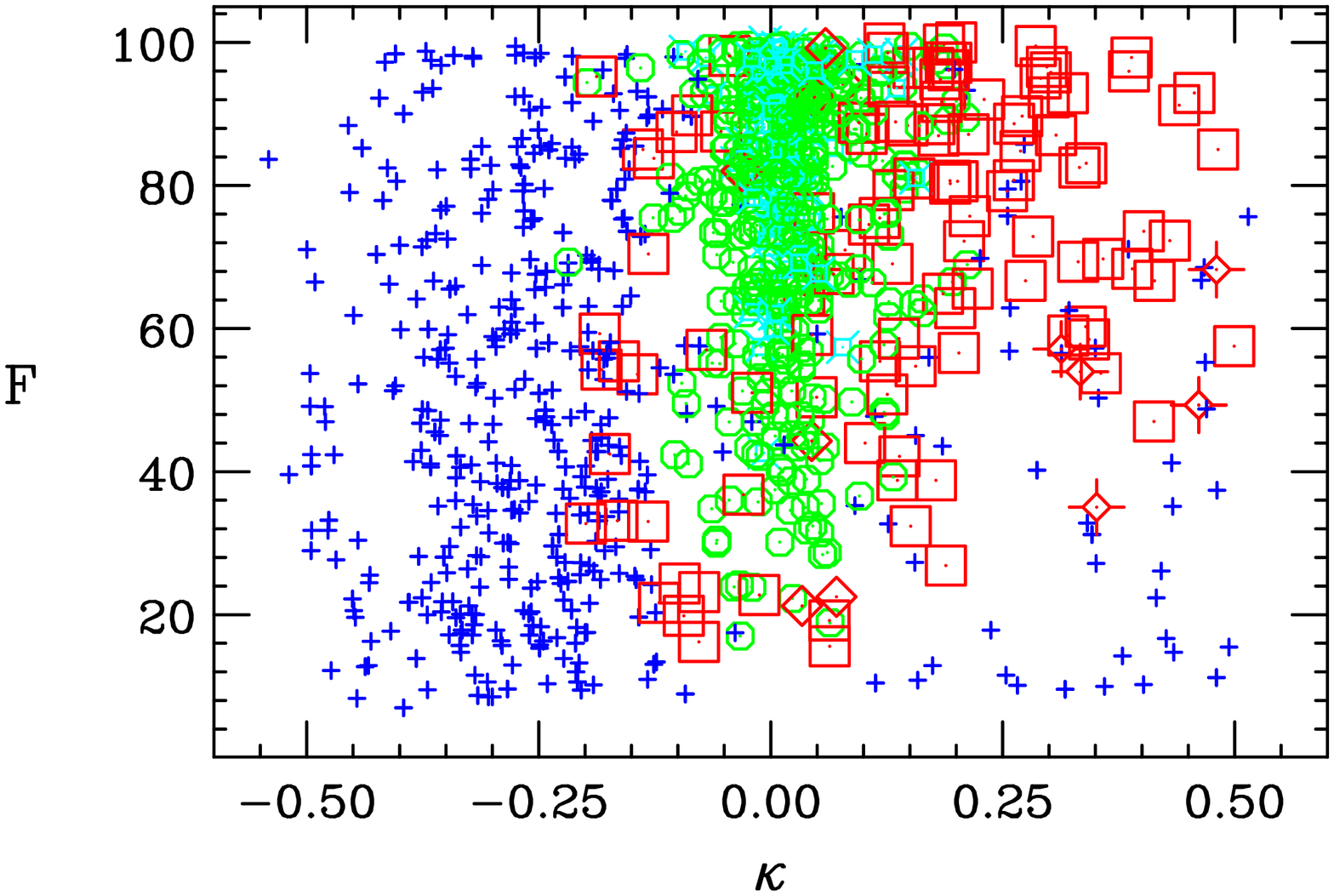}}
\vspace*{-.15in}
\caption{ For fixed $M_{1,2,3}(\mz)=100,200,300\gev$
and $\tanb=10$ we plot $F$ as a function of $\lam$ and
$\kap$. Notation and conventions as in Fig.~\ref{mixplots1}.}
\label{mixplots7}

\end{figure}
\begin{figure}[h!]
  \centerline{\includegraphics[width=2.4in,angle=90]{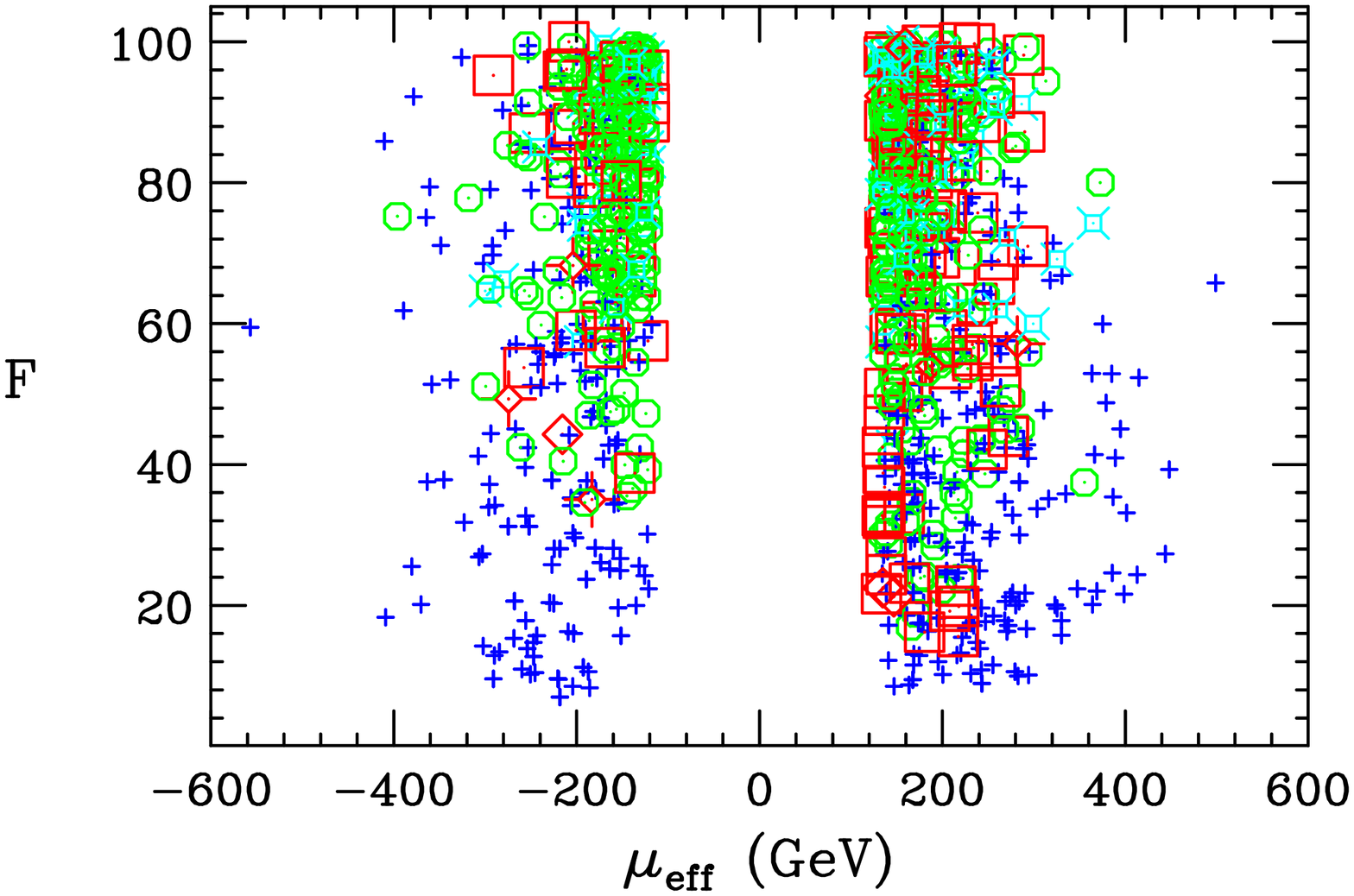}}
\vspace*{-.15in}
\caption{ For fixed $M_{1,2,3}(\mz)=100,200,300\gev$
and $\tanb=10$ we plot $F$ as a function of $\mueff$.
Notation and conventions as in Fig.~\ref{mixplots1}.}
\label{mixplots8}

\end{figure}

In Fig.~\ref{mixplots6}, we show $F$ as a function of
$\alam(\mgut)$ and $\akap(\mgut)$. There is considerable spread.
One noteworthy feature is that very small $F$ values can be achieved
for the (blue) $+$'s for $\alam(\mgut)$ and $\akap(\mgut)$ near zero.
The other noteworthy feature is that the MSSM-like mixed-Higgs
scenarios typically arise for substantial $\akap(\mgut)$.

Looking at Figs.~\ref{mixplots3}-\ref{mixplots6} in an overall sense,
we see that SUSY breaking Higgs, squark and mixing parameters should
be chosen at the GUT scale according to at least an approximate,
meaning $M_{SUSY}\lsim few\times 100\gev$, 'no-scale' model of SUSY
breaking in order to get the (blue) $+$ points that minimize fine
tuning, whereas this is not true for the mixed-Higgs scenarios.

Next, in Fig.~\ref{mixplots7}, we show $F$ as a function of $\lam$ and
$\kap$.  Note how the (blue) $+$'s with the very lowest $F$ values
populate a fairly distinct region from the mixed-Higgs scenario
points. In particular, the singlet mixed Higgs scenarios (green
circles and cyan starred-squares) populate a
region with $|\kap|<0.1$, whereas the (blue) $+$'s typically have
$|\kap|>0.1$. 

Finally, in Fig.~\ref{mixplots8}, we give $F$ as function of $\mueff$.
Obviously, small fine-tuning, whether in the mixed-Higgs scenarios or
in the non-tuned $\mai<2\mb$ scenarios, requires $\mueff$ between the lower bound of
about $120\gev$ allowed by LEP limits on the chargino mass and roughly
$250\gev$.  This would imply that charginos will be copiously produced
and probably easy to detect at the LHC and ILC, and probably reachable
in Tevatron late-stage running.

\section{Conclusions}

In this paper, we have explored the degree of fine-tuning associated
with mixed-Higgs scenarios, both in the MSSM and the NMSSM.  In the
MSSM, we have seen that, relative to the usual decoupled scenarios
with a lightest Higgs mass $\mh>114\gev$, mixed-Higgs scenarios allow a
reduction in the fine tuning, as measured by $F$ of Eq.~(\ref{fdef}),
of the GUT-scale model parameters in order to achieve correct EWSB.
The smallest $F$ achievable in the decoupled scenarios is $F\sim 30$,
while mixed-Higgs scenarios can be found with $F\sim 16$. Thus,
the mixed-Higgs MSSM scenarios give the smallest $F$ values among
those that are consistent with LEP limits. 

In the NMSSM, there are many parameter choices for which the lightest
Higgs has $\mhi>114\gev$, but, as in the MSSM case, the minimum $F$
values possible for such scenarios are large, $F\gsim 20$. (This,
however, is smaller than the minimum $F\sim 30$ achievable without
Higgs mixing in the MSSM scenarios with $\mh>114\gev$.) In the NMSSM,
further reduction in $F$ is possible in two distinct cases.  Scenarios
corresponding to the first are those with substantial Higgs mixing for
which $\mhi<114\gev$ is allowed by virtue of reduced $ZZ\hi$ coupling.
$F$ values as small as $\sim 16$ (6.5\% GUT-scale parameter tuning)
are possible in this first class of models.  Scenarios belonging to
the second class are those where $\mhi<114\gev$ and the $ZZ\hi$ has
full SM strength, but LEP constraints are satisfied because the
primary decay of the $\hi$ is $\hi\to \ai\ai\to 4\tau$ or $4~jets$.
Values of $F$ as small as $\sim 6$ (17\% GUT-scale parameter tuning,
which we regard as absence of fine tuning) are possible in this case.
This class of model is called the light-$\ai$ class. In a broad scan
over parameter space, light-$\ai$ models emerge more or less
immediately and automatically, whereas to find a significant number of
mixed-Higgs scenario with reasonably low $F$ requires highly focused
scans.

The mixed-Higgs scenarios in the NMSSM can be divided
into two classes: i) those in which the two doublet Higgs mix in close
analogy with the mixed-Higgs MSSM scenarios; and ii) those in which
there is substantial mixing of the doublet Higgses with the singlet
Higgs. The former class arises when the singlet Higgs decouples from
the doublet Higgses and there are many common features with the MSSM
mixed-Higgs scenarios.  In both the MSSM and class-i) NMSSM mixed
Higgs scenarios, one finds the lowest $F$ values for $\mhi\sim\mai\sim
100\gev$ and $\mhii\sim 120\gev$.  The corresponding
soft-SUSY-breaking parameters are essentially the same as well; in
particular, $A_t\sim -400\gev$ and $\mstopbar\sim 300\gev$. In the
class-ii) NMSSM Higgs scenarios with large singlet mixing, a large
range of $\mhi$ values is possible, but those with the smallest $F\sim
16$ values have $\mhi\sim 100\gev$ and $\mhii\sim 120\gev$, as
above, but a large range of possible $\mai$ values; $A_t\sim -400$ and
$\mstopbar\sim 300\gev$ are again needed. 

The light-$\ai$ NMSSM scenarios are quite different in nature. Minimal
$F$ is achieved when the $\hi$ is
very SM-like  and has mass $\mhi\sim 100\gev$. For
these scenarios, a large range of $\mhii$ is possible, beginning at
$\mhii\sim 150\gev$ and on up. The minimal $F$
values are achieved for  $A_t\sim -250\gev$ and somewhat smaller
$\mstopbar$. GUT-scale Higgs and soft-SUSY-breaking parameters are 
relatively close to those expected for no-scale SUSY breaking.
We have noted that the light-$\ai$ scenarios  also provide a natural
explanation of two crucial experimental observations: 1) the  $\hi$,
having $C_V^2(\hi)\sim 1$ and
$\mhi\sim 100\gev$,  provides a natural explanation of the precision electroweak
constraints; and 2) a value of $\br(\hi\to b\anti b) \sim 0.1$ 
is typical and
yields a good description of the LEP excess in the $\epem \to Z+b\anti
b$ channel at $M_{b\anti b}\sim 98\gev$.

The above can be contrasted with the mixed-Higgs scenarios.  While
these scenarios can also explain the excess of $Z+b\anti b$ events,
the required values of $\xi^2(Z+b's)\sim 0.1$ and $\mhi\sim 100\gev$
are only obtained if $F\gsim 30$ --- the lower $F\sim 16$
mixed-Higgs scenarios typically have $\xi^2(Z+b's)\sim C_V^2(\hi) \sim
0.2$ (see Figs.~\ref{mixplots2} and \ref{xisqh1}) and are thus only
barely consistent with LEP limits. In comparison, light-$\ai$
scenarios with the lowest $F\sim 6$ values always have $\mhi\sim
100\gev$ and a large fraction of these have $\xi^2(Z+b's)\sim 0.1$
(see Fig.~29 of Ref.~\cite{Dermisek:2007yt}).  In addition, a
mixed-Higgs scenario with $\mhi\sim 100\gev$ and $C_V^2(\hi)\sim 0.1$,
to explain the LEP excess, always has an $\hii$ with $\mhii> 114\gev$
and $C_V^2(\hii)\sim 0.9$, which combination does not yield nearly as
good agreement with precision electroweak data as the light-$\ai$
scenarios that always have $\mhi\sim 100\gev$ along with
$C_V^2(\hi)\sim 1$.

All cases discussed above have differences that will be clear once
experimental data for the Higgs sector become available. One important
test of the models will be consistency between the Higgs sector and
the stop sector. In particular, large mixing in the stop sector plays
a crucial role in the naturalness of EWSB in the MSSM. However, it is
highly nontrivial to measure the mixing at colliders. Some methods to
shed light on the mixing in the stop sector have been recently
discussed in Refs.~\cite{Dermisek:2007fi,Perelstein:2007nx}, but more
work in this direction is certainly desirable.

While it is true that the above MSSM and NMSSM scenarios can have
smaller fine tuning (as we define it) than those yielding a light
Higgs with mass above $114\gev$, these lower-$F$ scenarios always
require some additional restrictions (tuning) on other parameters,
\eg\ $m_{H_d}$ and $B_\mu$ in the MSSM case and similar parameters in
the NMSSM. This is to be contrasted with the fact that these same
parameters are not particularly constrained in the cases where the
lightest CP-even Higgs is SM-like. As a result of the parameter
correlations required to obtain the mixed-Higgs scenarios with low $F$
being significant, it might be very difficult to come up
with models in which low-$F$ mixed-Higgs scenarios are generic. In
this respect, the light-$\ai$ NMSSM models may have an edge by virtue
of the fact that a light $\ai$ is quite naturally obtained as a result
of a small breaking of the $U(1)_R$ symmetry limit of
$\akap(\mz)=\alam(\mz)=0$ via evolution from small $\akap$ and $\alam$
values at the GUT scale (see Refs.~\cite{Dobrescu:2000jt} and
\cite{Dermisek:2006wr}). Typical values for $\akap(\mgut)$ and
$\alam(\mgut)$ in the untuned $F\sim 6$ light-$\ai$ scenarios are shown in
Fig.~\ref{mixplots6}. In addition, such scenarios appear frequently for values
of the GUT-scale Higgs masses-squared and a GUT-scale $A_t$ value that
are all close to zero~(see Figs.~\ref{mixplots3} and \ref{mixplots4} and
further figures in Ref.~\cite{Dermisek:2007yt}). Thus, the GUT-scale
values for $\akap$, $\alam$, $\mhusq$, $\mhdsq$, $\mssq$ and $A_t$ are
quite consistent with an approximate no-scale model of SUSY breaking
in the case of light-$\ai$ scenarios with minimal fine tuning.

\begin{acknowledgments}
  RD would like to thank N. Maekawa for helpful discussions. This work
  was supported by the U.S. Department of Energy under grants
  DE-FG02-90ER40542 and DE-FG03-91ER40674.  JFG also receives support
  from the U.C. Davis HEFTI program. JFG thanks the Aspen Center for
  Physics where part of this work was performed.
\end{acknowledgments}

\end{document}